\documentclass[aps, twocolumn, pre, floatfix, superscriptaddress,longbibliography]{revtex4-2}

\usepackage{graphicx}
\usepackage{bm}
\usepackage{amsmath,amssymb,amsfonts,amsthm}
\usepackage{physics}
\usepackage{color}
\usepackage{bbold}

\definecolor{darkblue}{rgb}{0,0,.65}
\definecolor{darkgreen}{rgb}{0.3,0.6,0.3}
\definecolor{cyan1}{rgb}{0.0, 0.6, 0.6}
\usepackage[%
pdfstartview=FitH,%
breaklinks=true,%
bookmarks=true,%
colorlinks=true,%
anchorcolor=black,%
citecolor=blue,
filecolor=black,%
menucolor=black,%
urlcolor=darkblue,%
linkcolor=blue,%
]{hyperref}

\providecommand{\innerprod}[2]{\ensuremath{ \left\langle #1 | #2 \right\rangle}}
\providecommand{\outerprod}[2]{\ensuremath{| #1 \rangle \langle #2 | }}
\providecommand{\ket}[1]{\ensuremath{\left|{#1}\right.\rangle}}
\providecommand{\bra}[1]{\ensuremath{\langle\left.{#1}\right|}}

\DeclareMathOperator*{\argmin}{argmin}

\graphicspath{{.}{./CFigures/}}

\begin{document}
	
	\title{Assigning Temperatures to Eigenstates}
	
	\author{Phillip C. Burke}
	
	\affiliation{Department of Theoretical Physics, Maynooth University, Maynooth, Kildare, Ireland}
	
	\author{Goran Nakerst}
	
	\affiliation{Institut f\"ur Theoretische Physik, Technische Universit\"at Dresden, 01062 Dresden, Germany}
	
	\affiliation{Department of Theoretical Physics, Maynooth University, Maynooth, Kildare, Ireland}
	
	\author{Masudul Haque}
	
	\affiliation{Institut f\"ur Theoretische Physik, Technische Universit\"at Dresden, 01062 Dresden, Germany}
	
	\affiliation{Department of Theoretical Physics, Maynooth University, Maynooth, Kildare, Ireland}
	
	\affiliation{Max-Planck Institute for the Physics of Complex Systems, Dresden, Germany}
	
	\date{\today}
	
	\begin{abstract}
		
		In the study of thermalization in finite isolated quantum systems, an inescapable issue is the definition of temperature. 
		We examine and compare different possible ways of assigning temperatures to energies or equivalently to eigenstates in such systems. 
		A commonly used assignment of temperature in the context of thermalization is based on the canonical energy-temperature relationship, which depends only on energy eigenvalues and not on the structure of eigenstates. 
		For eigenstates, we consider defining temperature by minimizing the distance between (full or reduced) eigenstate density matrices and canonical density matrices. 
		We show that for full eigenstates, the minimizing temperature depends on the distance measure chosen and matches the canonical temperature for the trace distance; however, the two matrices are not close. 
		With reduced density matrices, the minimizing temperature has fluctuations that scale with subsystem and system size but appears to be independent of distance measure. In particular limits, the two matrices become equivalent while the temperature tends to the canonical temperature.
		
	\end{abstract}
	
	\maketitle
	
	
	\section{Introduction}\label{sec:Introduction}
	
	In recent years, there has been significant interest in reconciling statistical mechanics to the quantum dynamics of isolated many-body systems. 
	This endeavor invariably requires a correspondence between energy, a quantity well-defined in quantum mechanics, and temperature, which is necessary for a statistical-mechanical description.  
	The eigenstate thermalization hypothesis (ETH) \cite{Deutsch_PhysRevA.43.2046, Srednicki_PhysRevE.50.888, Srednicki_1996,
		Srednicki_1999, Reimann_NJP2015, Deutsch_RepProgPhys2018,
		Mori_Ikeda_Ueda_thermalizationreview_JPB2018,rigol_thermalization_2008,
		Rigol_Srednicki_2012, Alessio_Rigol_AdvPhys2016}, a cornerstone of this field, posits that each eigenstate contains information relevant to thermalization. 
	Thus, a natural question is how to assign temperatures to each eigenstate based on information encoded in the eigenstates. 
	In this work, we examine possible ways of doing so.
	
	The standard definition of temperature in statistical mechanics is given by the
	inverse of the derivative of entropy with respect to energy
	\cite{Reif_StatPhys,kardar_statphysp_2007}. For an isolated quantum system, the entropy at energy
	$E$ is defined as the logarithm of the number of microstates (i.e., eigenstates) with energy $E$,
	or energy in a window around $E$. In finite systems, obtaining this entropy generally requires
	approximating the density of states.
	
	Within the context of thermalization in finite isolated quantum systems, it is more common to use the
	canonical temperature-energy relationship to extract temperature from the eigenvalues of the system
	Hamiltonian.  The canonical temperature $T_{C}=1/\beta_{C}$ can be obtained for any energy $E$ by inverting the canonical equation  
	%
	%
	\begin{equation}\label{eq:canoncial_temp}
		E = \langle H \rangle \ = \ \frac{\tr( e^{-\beta_{C} H}H)}{\tr( e^{-\beta_{C} H})} \ = \ \frac{ \sum_{j}
			e^{-\beta_{C} E_{j}}E_{j} }{\sum_{j} e^{-\beta_{C} E_{j}}} , 
	\end{equation} 
	where $E_j$ are the eigenvalues of the system Hamiltonian $H$.  This relationship originates in
	statistical mechanics from the context of a system with a bath, but is widely used in the study of
	the thermalization of isolated (bath-less) quantum systems to obtain an energy-temperature
	correspondence \cite{rigol_thermalization_2008, Rigol_PRL2009, Rigol_PRA2009, RigolSantos_PRA10,
		Rigol_Srednicki_2012, Santos_Polkovnikov_Rigol_PRE2012, Neunhahn_Marquardt_PRE2012,
		SorgVidmarHeidrichMeisner_PRA14, NandkishoreHuse_AnnuRev2015,
		Essler_2016_JoSM,Seki_Yunoki_PRResearch2020, Greinergroup_thermalization_Science2016,
		Alessio_Rigol_AdvPhys2016, Garrison_PhysRevX.8.021026,Santos_Rigol_onset_PRE2010,
		Roux_PRA2010_quantumquenches,FratusSrednicki_PRE2015, Noh_PRE2021,Rigol_Srednicki_PRL2013}. 
	In the large-size limit, the canonical temperature is, of course, equivalent to that obtained by
	differentiating the entropy.
	
	Curiously, both of these definitions rely only on the energy eigenvalues, 
	making no reference to the physics of the eigenstates.
	Therefore, it is of obvious interest to compare the temperatures obtained from eigenstates 
	($\beta_{E}$ and $\beta_{S}$, introduced below) with an eigenvalue-based definition.  In this work,
	we introduce ways of obtaining temperatures from eigenstates and then compare them to the canonical
	temperature, $\beta_{C}$, widely used in the thermalization literature. 
	

	If an eigenstate $\ket{E_n}$ of a many-body system `knows' the temperature corresponding to
	its energy $E_n$, then one might na{\"i}vely expect that $\rho=\outerproduct{E_n}{E_n}$
	should be closest to the canonical density matrix (DM) $\rho_{C}=Z^{-1}e^{-\beta H}$ for
	that value of the inverse temperature $\beta$.  (Here $Z=\tr e^{-\beta H}$.)  Thus,
	minimizing the distance $d(\rho,\rho_C)$ between these two DMs as a function of $\beta$ is
	one way of assigning a temperature to $E_n$.  We refer to this optimal $\beta$ as the
	`eigenstate temperature' $\beta_{E}$.  As $\outerproduct{E_n}{E_n}$ is the limit of the
	microcanonical DM for an ultra-narrow energy window, this idea is also related to the
	equivalence of statistical ensembles \cite{Gurarie_AmJPhys2007,Tasaki_equivalence_JSP2018}
	--- this definition of temperature minimizes the distance between microcanonical and
	canonical DMs.
	
	It is admittedly over-ambitious to expect the complete eigenstate DM $\rho$ to resemble a
    Gibbs thermal state $\rho_{C}$, since the first is a pure state and the second is a mixed
    state.  The two density matrices cannot be expected to be `close', as we will illustrate in
    Section \ref{sec:eigenstate_temperature}.  In real-time dynamics, the common inquiry is
    whether a local sub-region, rather than the whole system, approaches a thermal state
    \cite{Cramer_Eisert_Osborne_PRL2008, Linden_Popescu_Short_Winter_PRE2009,
      Rajagopal_oscillators_PRE2009, Short_NJP2011, Gogolin_Eisert_PRL2011,
      Riera_Gogolin_Eisert_PRL2012, Brandao_Horodecki_equilibration_PRE2012,
      Genway_Ho_Lee_PRA2012, DePalma_Cramer_necesssityETH_PRL2015,
      Eisert2015_Gogolin_review2015, Greinergroup_thermalization_Science2016,
      Farrelly_Brandao_Cramer_PRL2017, Seki_Yunoki_PRResearch2020}. The intuition is that the
    rest of the system acts as an effective bath, even if the textbook properties of a bath
    (weak coupling, no memory) are not satisfied. Accordingly, ETH is often formulated in terms
    of local observables or a spatial fraction of the system
    \cite{Beugeling_Moessner_Haque_PRE2014, Eisert2015_Gogolin_review2015,
      Beugeling_Moessner_Haque_PRE2015, Muller2015, DePalma_Cramer_necesssityETH_PRL2015,
      Greinergroup_thermalization_Science2016, Mondaini_Rigol_2DIsing_PRE2017,
      Dymarsky_Lashkari_Liu_PRE2018, Garrison_PhysRevX.8.021026, Dymarsky_PRB2019}, and similar
    ideas appear in the approach known as canonical typicality \cite{Tasaki_PRL1998,
      Goldstein_PhysRevLett.96.050403, Popescu2006, Reimann_typicality_PRL2007,
      Linden_Popescu_Short_Winter_PRE2009, Short_NJP2011, Santos_Polkovnikov_Rigol_PRE2012,
      Muller2015, Mori_Ikeda_Ueda_thermalizationreview_JPB2018}. Thus, one expects for
    thermalizing systems that, if the system is partitioned spatially into $A$ and $B$, with $A$
    smaller, then the reduced DM of subsystem $A$ for an eigenstate, $\rho^{A} = \tr_{B}\rho$,
    should approximate the reduced canonical DM, $\rho_{C}^{A} = \tr_{B}\rho_{C}$
    \cite{Muller2015, Garrison_PhysRevX.8.021026, Dymarsky_Lashkari_Liu_PRE2018}.  Inverting
    this expectation, we obtain another way of assigning temperatures to eigenstates --- use the
    value of $\beta$ which minimizes the distance $d(\rho^{A},\rho_{C}^{A})$.  We call this the
    `subsystem temperature' $\beta_{S}$.

	We find that $\beta_{E}$, which minimizes the distance between canonical DMs $\rho_{C}$ and
	eigenstate (or microcanonical) DMs $\rho$, depends on the distance measure employed.  Using
	distances based on the Schatten $p$-norm \cite{book_Laub_Matrix_2005, book_Horn_Johnson_Matrix_2013,
		book_Zhan_Matrix_2013, book_Bhatia_Matrix_2013}, we show analytically that the minimizing
	temperature $\beta_{E}$ is equal to $p^{-1}$ times the canonical temperature $\beta_{C}$.  Thus,
	only the trace distance ($p=1$) gives meaningful physical results; even the well-known
	Hilbert-Schmidt norm ($p=2$) would provide a temperature that deviates by a factor of two!  Although $\beta_{E}$ aligns with $\beta_{C}$ for $p=1$, the two DMs are never close, i.e., even the minimum distance is large.
	
	The subsystem temperature $\beta_{S}$ appears numerically to be broadly independent of $p$ and is seen to match the canonical temperature $\beta_{C}$ only approximately at finite sizes. 
	Thus for finite systems, the reduced DMs of pure eigenstates can be closer to thermal states at temperatures other than the canonical temperature.  
	The correspondence is shown to improve in the limit where the size of the subsystem complement ($B$) is large, but not necessarily in other ways
	of taking the large-size limit.
	
	The paper is laid out as follows. In Section \ref{sec:prelims}, we outline the distance measures used to quantify how close two density matrices are and introduce the many-body quantum systems that we will numerically investigate. Following this, we present our results for the eigenstate and subsystem temperatures in Sections \ref{sec:eigenstate_temperature} and \ref{sec:Subsystem_Temperature}, respectively. In Section \ref{sec:Alternate_formulations}, we outline alternative choices that could have been used in our investigations. Then, in Section \ref{sec:non_chaotic}, we investigate the deviation of the subsystem temperature from the canonical temperature as a system approaches integrability. Finally, in Section \ref{sec:Discussions}, we summarize our findings and discuss their relation to existing work. In addition, we outline open questions that remain. 
	
	\section{Preliminaries}\label{sec:prelims}
	
	Here we first define an appropriate distance measure between density matrices. This distance measure
	is to be used in our temperature definitions. Following this, we describe the many-body quantum
	systems used in numerical calculations. For each system, we provide the relevant quantum
	Hamiltonian.
	
	\subsection{Distance Measures}\label{sec:dist_measures}
	
	To quantify the distance between two DMs, we use the Schatten $p$-distance, the norm of the difference between the two normalized matrices
	\begin{equation}\label{eq:p_dist_def}
		d_p(\rho, \sigma) = \left\| \ \frac{\rho}{\|\rho\|_p} - \frac{\sigma}{\|\sigma\|_p}\ \right\|_p,
	\end{equation}
	with the Schatten $p$-norm 
	given by 
	\begin{equation}
		\|A\|_p=\tr\left(|A|^p\right)^{1/p}= \left(\sum_n |s_n|^p\right)^{1/p}, 
	\end{equation}
	for a Hermitian matrix $A$ and $1\le p < \infty$.  Here $s_n$ are the singular values of $A$, and
	$|A| = \sqrt{A^{\dagger} A}$.
	This class of distances includes commonly used measures of distance between DMs, such as the trace
	distance \cite{book_Barnett_QuantInfo_2009, nielsen_chuang_2010} and the Hilbert-Schmidt (or
	Frobenius) distance \cite{Baltz_dist_IOP_1990,Knoll_Orlowski_dist_PRA1995,
		Hillery_dist_PRA1996,Dodonov_Wunsche_dist_IOP1999,
		Dodonov_Manko_Wunsche_dist_JoOptics2000,Wunsche_Dodonov_Manko_ProgPhys2001,
		Dodonov_Reno_dist_PhysLettA2003,Scutaru_dist_PRA2004,
		Genoni_Paris_quant_PRA2010,Roga_Illuminati_IOP2016,
		Bartkiewics_Lemr_PRA2019,Coles_etal_TraceDist_vs_HSdist_PRA2019,
		Lemr_measure_PRL2019,Kumar_random_PRA2020,Park_Hyunchul_quant_PRA2021}.
	The range of $d_p$ is  $[0,2]$.  
	
	The main body of this paper are based on the Schatten $p$-distances.  In Section \ref{sec:FidBure},
	we will examine briefly how our results are affected if one uses instead the Bures distance
	\cite{book_Barnett_QuantInfo_2009, nielsen_chuang_2010}.
	
	
	\subsection{Many-body systems}\label{sec:manybody_systems}
	
	To ensure that the presented results hold generally for chaotic (thermalizing) many-body
	Hamiltonians with local interactions, we will provide numerical results for three different 1D, and
	a 2D, non spin conserving, chaotic models. For all systems, we consider a spin-$\frac{1}{2}$ lattice
	of $L$ sites with open boundary conditions.

	The first model is the quantum Ising model, with transverse and longitudinal magnetic fields on
	every site. The transverse and longitudinal fields have strength $h_x$ and $h_z$ respectively. To
	remove symmetries of the model, we swap the $x$ and $z$ field strength between the first two
	sites. The chaotic Ising Hamiltonian is then
	\begin{align}\label{eq:Ham_Ising}
		H_I = \sum_{j=1}^{L-1}S_j^{z}S_{j+1}^{z} + \sum_{j=1}^{L}&(h_{x}(1-\delta_{j,1} ) S^{x}_{j} +h_{z}(1-\delta_{j,2})S^{z}_{j} )\notag \\
		&+h_{z}S^{x}_{1} + h_{x}S^{z}_{2}.
	\end{align}
	
	The second model is the \textit{XXZ}-chain with staggered transverse and longitudinal magnetic fields along the even and odd sites respectively. In addition, we break the staggered pattern at the start of the chain by inserting $x$ and $z$ fields on the first and second sites respectively to remove any symmetry. The staggered \textit{XXZ}-chain Hamiltonian is then
	\begin{align}\label{eq:Ham_Stag}
		H_S =& \sum_{j=1}^{L-1}\left( S^{x}_{j}S^{x}_{j+1} + S^{y}_{j}S^{y}_{j+1}  +\Delta S_j^{z}S_{j+1}^{z} \right) \notag\\
		&+ \sum_{\text{even}}h_{x}S^{x}_{j} +\sum_{\text{odd}}h_{z}S^{z}_{j} +h_{x}S^{x}_{1} + h_{z}S^{z}_{2}.
	\end{align}
	
	The last 1D model we used is the \textit{XXZ}-chain with disordered transverse and longitudinal magnetic fields on every site. In this case, rather than $h_z$ and $h_x$ being uniform across the sites, the on-site strengths $h_j$, $h'_j$, are chosen from a uniform distribution $[-W,W]$. The disordered \textit{XXZ}-chain Hamiltonian is then
	\begin{align}
		H_D = \sum_{j=1}^{L-1}&\left( S^{x}_{j}S^{x}_{j+1} + S^{y}_{j}S^{y}_{j+1}  +\Delta S_j^{z}S_{j+1}^{z} \right) \notag\\
		&+ \sum_{j=1}^{L}(h_jS^z_j +h'_jS^x_j ).
		\label{eq:Ham_dis}
	\end{align}
	For all three 1D models, appropriate parameters were chosen to ensure chaotic level spacing statistics. Namely, $h_z = 0.5$, $h_x = 0.75$ for the Ising model, $h_z = h_x = 0.5$ for the staggered field model, and $W=0.25$ for the disordered field model.
	
	Finally, the 2D model we use is a square lattice, with \textit{XXZ}-like connections between neighboring spins $\langle j,k\rangle$. In addition, transverse magnetic fields are placed on the sites $j_a$ in one of the sub-lattices available within the bipartite square lattice, in order to break total spin conservation. The square lattice Hamiltonian is given by
	\begin{align}
		H_{sq} = \sum_{\langle j,k\rangle}&\left[J_{jk}\left( S^{x}_{j}S^{x}_{k} + S^{y}_{j}S^{y}_{k}\right)  +\Delta_{jk} S_j^{z}S_{k}^{z} \right] \notag\\
		&+ \sum_{j_a}h_{x}S^{x}_{j_a}.
		\label{eq:Ham_Sq}
	\end{align}
	To ensure chaotic level spacing statistics, the parameters $J_{jk}$ and $\Delta_{jk}$ are drawn randomly from the uniform distribution $[0,2]$ and $[0,1]$ respectively. This choice of parameters ensures any symmetries of the lattice are broken.
	
	For 1D systems, the $A$ subsystem is taken to be the leftmost $L_A$ sites of the $L$-site chains.
	In the 2D square lattice, the $A$ subsystem is taken to be the first $L_A$ consecutive sites, starting from a corner of the square and following either a row or column. When this model is used, illustrations of the lattice geometry are provided.  
	For simplicity, we choose systems whose underlying Hilbert space $\mathcal{H}$ has a tensor product structure
	$\mathcal{H}=\mathcal{H}_A\otimes\mathcal{H}_B$.  This is the case for spin and fermionic systems,
	where total spin and particle number respectively are not conserved.  Then the full Hamiltonian can
	be written as $H = H_A \otimes \mathbb{1}_{D_B} + \mathbb{1}_{D_A} \otimes H_B + H_{AB}$, where $H_A$ and $H_B$ only
	act on $A$ and $B$ respectively, and $H_{AB}$ is the interaction between the two.  The Hilbert space
	dimensions of $A$, $B$ and the total system are $D_A$, $D_B$ and $D = D_{A}D_{B}$ respectively.

	
	\section{Eigenstate Temperature}\label{sec:eigenstate_temperature}
	
	Here we discuss the eigenstate temperature, which we have defined as
	\begin{equation}\label{eq:estate_temp}
		\beta_{E} = \displaystyle\argmin_{\beta} d_p\left(\rho,\rho_C\right).
	\end{equation}
	Here, $\rho$ is an eigenstate density matrix, while $\rho_{C}$ is a canonical density matrix. We first present analytical results that are general to all Hermitian systems. In addition, we provide numerical results that illustrate these analytical results. Following this, we consider a variation of the eigenstate temperature. In particular, we consider a density matrix consisting of an equally weighted sum of eigenstates from a finite energy window, i.e., a microcanonical density matrix. Finally, we provide the full derivation of the analytical results presented.
	
	\subsection{Main Results}
	
	In order to determine the value of $\beta_{E}$,  
	we express the two density matrices in the basis for which they are simultaneously diagonalized, and set to zero the
	derivative of $d_p(\rho,\rho_{C})$ with respect to $\beta$. 
	The full derivation of the minimum can be found in Section \ref{sec:analytics}, the main result of which is that the minimum is precisely when
	\begin{equation}
		E_n = \frac{\tr(H e^{-p\beta H})}{\tr(e^{-p\beta H})}.
		\label{eq:mc_temp_result}
	\end{equation}
	Thus, comparing with the definition \eqref{eq:canoncial_temp} of the canonical temperature,
	\begin{equation}
		\beta_{E} = \frac{\beta_{C}}{p}. 
	\end{equation}
	The eigenstate and canonical temperatures coincide for $p=1$, while they differ by a factor of $p$
	for $p>1$.  This result is purely mathematical and holds for an arbitrary Hermitian matrix $H$,
	irrespective of whether $H$ has the interpretation of a many-body Hamiltonian, e.g., even for a
	random matrix, see results in Appendix \ref{sec:Random}.

	\begin{figure}
		\includegraphics*[width=\linewidth]{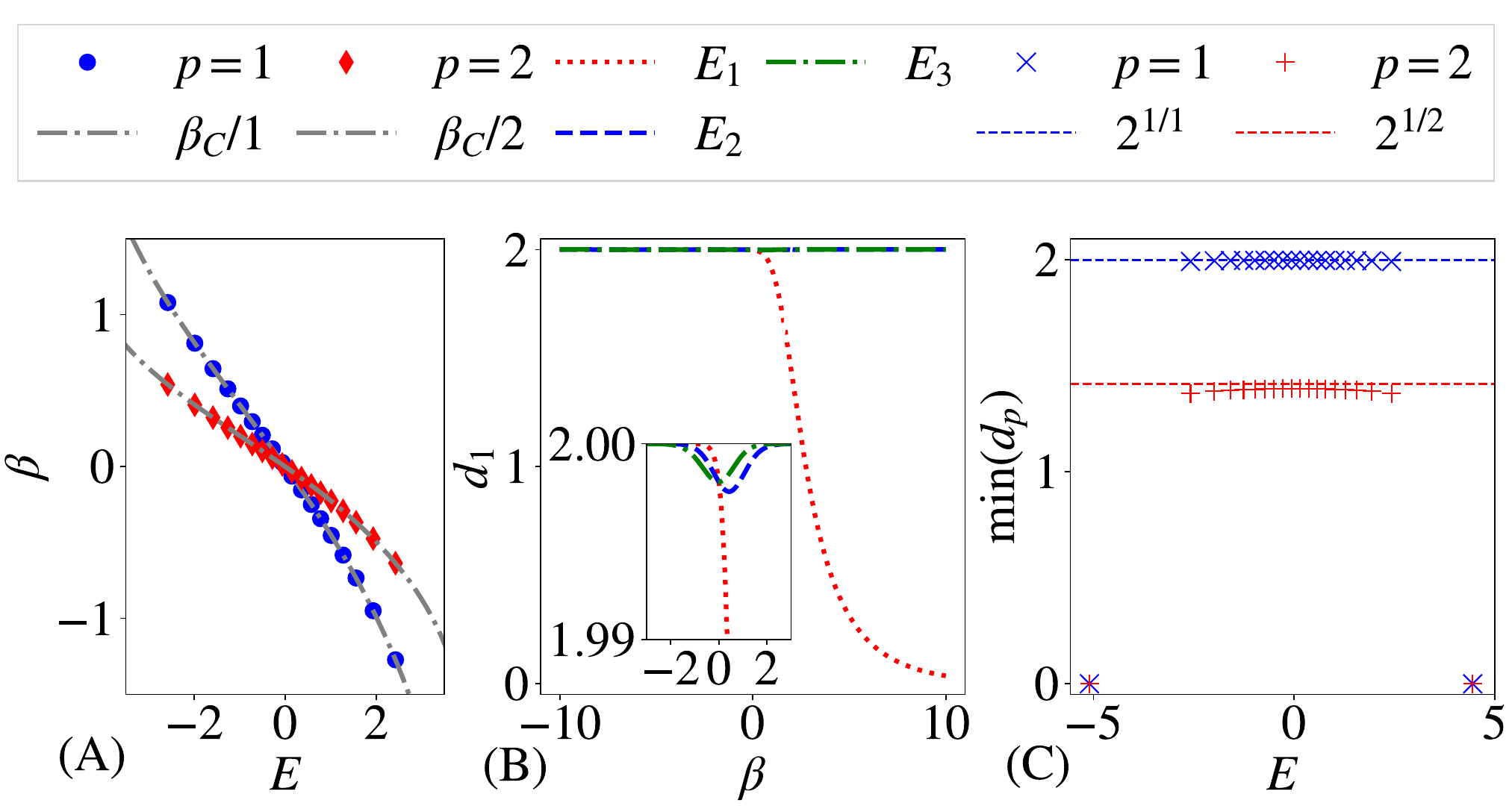}
		\caption{Eigenstate temperature  results for staggered field \textit{XXZ}-chain: $h_x = h_z = 0.5$, $\Delta=0.95$, $L=10$.
			\textbf{(A)}
			$\beta_E$ against energy, for 20 eigenstates which are equally spaced in energy across the spectrum, with curves showing $\beta_C/p$. (Highest/lowest state not visible.)  
			\textbf{(B)} $d_1(\rho,\rho_{C})$ vs $\beta$ curve for ground state ($E_1$), mid-spectrum
			state ($E_3$), and $E_2$ in between the two.  
			\textbf{(C)} The minimum of $d_p(\rho,\rho_{C})$ plotted against
			eigenenergy, for the same
			eigenstates used in (A). 
		}
		\label{fig:estateT_EvsB}
	\end{figure}

	Fig.~\ref{fig:estateT_EvsB} illustrates this relation $\beta_{C} = p\beta_{E}$ (A) and the
	behavior of the distance $d_p$  (B,C), for the staggered field \textit{XXZ}-chain.

	The result $\beta_{E} = \beta_{C}$ (for $p=1$) does not imply that eigenstate DMs
	$\rho=\outerproduct{E_n}{E_n}$ closely resemble canonical states $\rho_{C}=Z^{-1}e^{-\beta H}$.
	
	We are comparing a pure state to a highly mixed state, i.e., a projection operator (a rank-1
	operator) $\rho$ to a full-rank operator $\rho_{C}$.  So, even the smallest distance between them
	(at $\beta=\beta_{C}$) is close to the maximum. The smallest $p$-distance is in general close to
	$2^{1/p}$, an analytical result derived in the following Section \ref{sec:analytics}.
	The minimum is thus very close to the maximum for most eigenstates, as shown in Fig.~\ref{fig:estateT_EvsB}(B,C).
	The highest/lowest eigenstates are exceptions. 
	
	
	\subsection{Finite Window Eigenstate Temperature}\label{sec:Microcanonical_DM}
	
	Instead of the eigenstate DM, $\rho = \outerprod{E_n}{E_n}$, one could use the microcanonical
	DM, 
	\begin{equation}\label{eq:microcan_state}
		\rho_{MC} = \frac{1}{\mathcal{N}} \sum_{E_j\text{ in } \Delta E} \ket{E_j}\bra{E_j},
	\end{equation}
	where $\Delta{E}$ is an energy window containing $E_n$, and $\mathcal{N}$ is the number of states in the window. This might be considered more physical, as we are now comparing two mixed states.
	
	\begin{figure}
		\includegraphics[width=\linewidth]{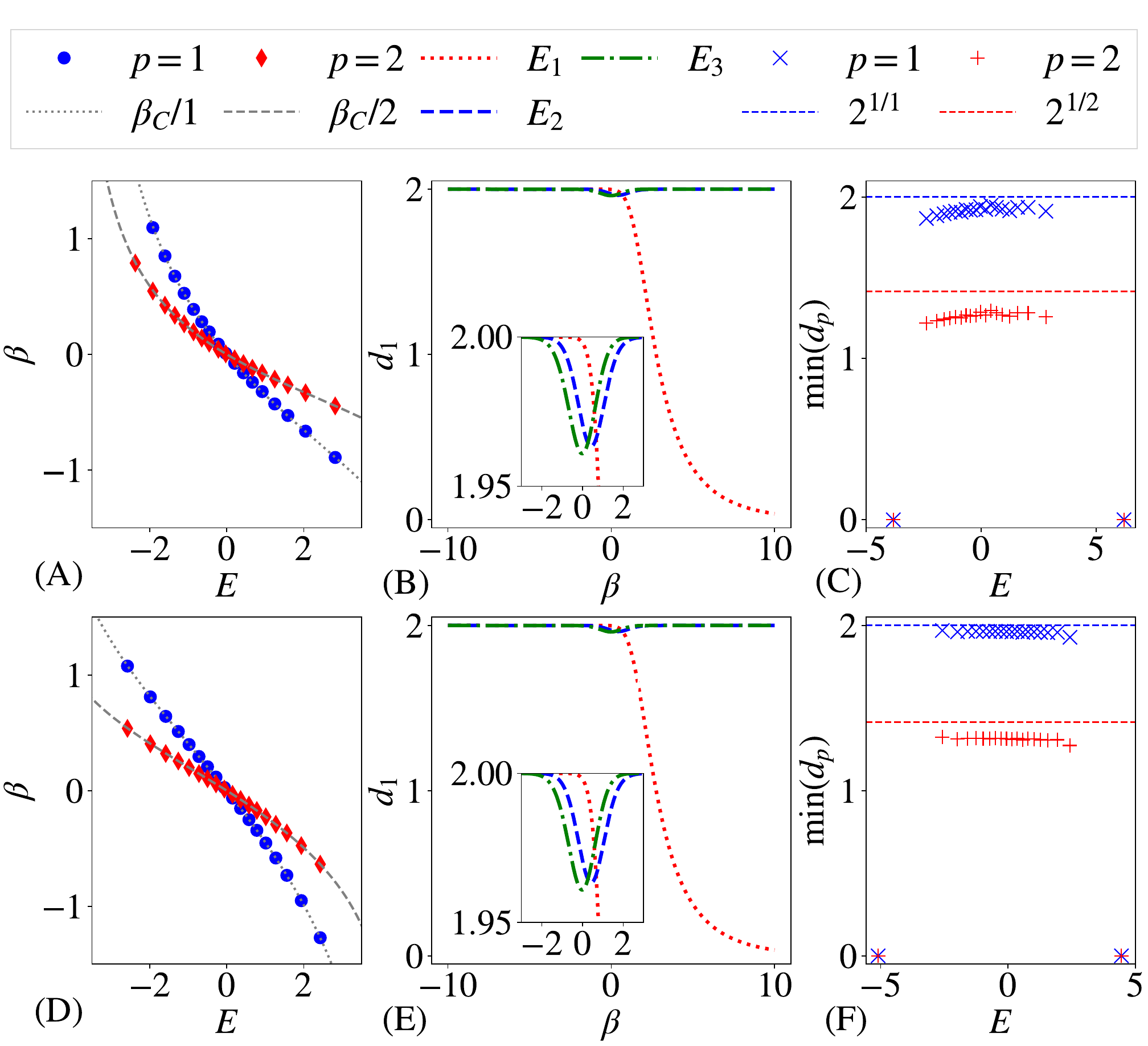}
		
		\caption{Finite window eigenstate temperature results for spin chains with $L=10$, namely:
			\textbf{(A)-(C)} Chaotic Ising Model with $h_z = 0.5$, $h_x=0.75$ and $\Delta E \sim 0.059$. 
			\textbf{(D)-(F)} Staggered field \textit{XXZ}-chain with $\Delta=0.95$, $J=1$, $h_x = h_z = 0.5$, and $\Delta E \sim 0.0757$.
			\textbf{(A,D)} $\beta_{MC}$ against energy, for 20 energy windows which are equally spaced in energy across the spectrum, with curves showing $\beta_C/p$. (Highest/lowest state not visible.)
			\textbf{(B,E)} $d_1(\rho_{MC},\rho_{C})$ vs $\beta$ curve for energy windows: $E_1$ near the ground state, $E_2$ in the middle of spectrum, and $E_2$ in between the two.
			\textbf{(C,F)} The minimum of $d_p(\rho_{MC},\rho_{C})$ plotted against
			energy, for the same energy windows used in (A,D).
		}
		\label{fig:microcan_EvsB}
	\end{figure}
	
	Here, we fix the energy window width, and allow each window to contain a different number of eigenstates. We want to compute the value of $\beta$ such that the distance $d_p(\rho_{MC},\rho_{C})$ is minimized. We label this minimizing value the finite window eigenstate temperature $\beta_{\Delta{E}}$. One can follow the same procedure as is detailed in Section \ref{sec:analytics} for the eigenstate temperature, and make the assumption that the energy $E_{MC} = \tr(H \rho_{MC})= 1/\mathcal{N}\sum_{ E_j \in \Delta E}E_j$ of the microcanonical state $\rho_{MC}$ is roughly $E_{MC} \approx E_j \in \Delta E$, which is valid if the energy interval is sufficiently small, and obtain the similar relation that $\beta_{C} \approx p\beta_{\Delta{E}}$.
	
	This result is illustrated numerically in Fig.$~$\ref{fig:microcan_EvsB}, in which we present results for the chaotic Ising model (A-C) and the staggered field \textit{XXZ}-chain (D-F). In (A,D), we plot $\beta_{\Delta{E}}$ that minimizes the Schatten $p$-distance for the given $p$, along with two canonical $\beta_{C}$ curves, versus energy. In (C,F), we plot the value of the minimum distance for the same energy slices as taken in the left figure. Finally, in (B,E) we plot the $d_1$ distance versus $\beta$ for three particular energy slices $E_1$, $E_2$ and $E_3$. The numerical results again illustrate the derived relation of $\beta_{C} = p\beta_{\Delta{E}}$ for the $p$-distance $d_p$ when taken between a microcanonical and canonical density matrix.
	
	
	\subsection{Derivation of analytical results}\label{sec:analytics}
	
	We wish to minimize the Schatten $p$-distance \eqref{eq:p_dist_def} between the canonical and eigenstate DMs, i.e., $d_{p}(\rho,\rho_{C})$.
	All Schatten $p$-norms of a matrix $\rho$ can be expressed in terms of the singular values $s_n$ of $\rho$
	\begin{equation}
		\|\rho\|_p = \left( \sum_n s_n^p \right)^{1/p}.
	\end{equation}
	In other words, the Schatten $p$-norm is the $l_p$ norm of the singular values. The singular values of a Hermitian matrix $\rho$ are the absolute values of the eigenvalues of $\rho$.
	The eigenstate density matrix $\rho$ and the canonical density matrix $\rho_C$ are jointly diagonalizable with respect to the eigenstate basis of $H$. The eigenvalues of the former are $1$ and $0$, while the eigenvalues of the latter are given by $e^{-\beta E_j}$, where $E_j$ are the eigenvalues of the Hamiltonian $H$. The Schatten norms are invariant under a basis transformation by definition, so the normed Schatten $p$-distance can be written as
	\begin{align}\label{eq:diff_p}
		d_p^p(\rho, e^{-\beta H})
		= \left| \frac{1}{\|\rho\|_p} - \frac{e^{-\beta E_n}}{\|e^{-\beta H}\|_p} \right|^p
		+ \sum_{ E_j \neq E_n} \frac{e^{-p\beta E_j}}{\|e^{-\beta H}\|_p^p}.
	\end{align}
	Now there are two results we wish to obtain, the value of $\beta$ for which \eqref{eq:diff_p} is minimized, and the value of that minimum. In \ref{sec:min} we obtain the surprising result of $\beta_{E} = \beta_{C}/p$, and in \ref{sec:min_val} we determine how the value of the minimum scales.
	
	\subsubsection{Minimization}\label{sec:min}
	
	To find the minimum of \eqref{eq:diff_p}, we differentiate the $p$-normed Schatten $p$-distance of $\rho$ and $\exp(-\beta H)$ and obtain	
	\begin{align}
		\frac{\partial}{\partial \beta} d_p^p(\rho,& e^{-\beta H})  
		=  
		-p \left( \frac{1}{\|\rho\|_p} - \frac{e^{-\beta E_n}}{\|e^{-\beta H}\|_p} \right)^{p-1} \nonumber \\
		&\times \frac{\partial}{\partial \beta} \frac{e^{-\beta E_n}}{\|e^{-\beta H}\|_p} 
		+ \sum_{E_{j}\neq E_{n}} \frac{\partial}{\partial \beta} \frac{e^{-p\beta E_j}}{\|e^{-\beta H}\|_p^p}.
		\label{eq:diff_deriv}
	\end{align}
	Then, we observe the two derivatives
	\begin{align}
		\frac{\partial}{\partial \beta} \frac{e^{-\beta E_n}}{\|e^{-\beta H}\|_p}
		&= \frac{e^{-\beta E_n}}{\|e^{-\beta H}\|_p} 
		\left( -E_n + \frac{\tr(He^{-p\beta H})}{\tr(e^{-p\beta H})} \right)\\
		\frac{\partial}{\partial \beta} \frac{e^{-p\beta E_j}}{\|e^{-\beta H}\|_p^p}
		&= \frac{pe^{-p\beta E_j}}{\|e^{-\beta H}\|_p^p} 
		\left( -E_j + \frac{\tr(He^{-p\beta H})}{\tr(e^{-p\beta H})} \right).
	\end{align}
	
	Now, \eqref{eq:diff_p} is minimal if and only if \eqref{eq:diff_deriv} is 0, which holds true if and only if
	
	\begin{align}
		0
		&= -p\left( \frac{1}{\|\rho\|_p} - \frac{e^{-\beta E_n}}{\|e^{-\beta H}\|_p} \right)^{p-1} \frac{e^{-\beta E_n}}{\|e^{-\beta H}\|_p} \nonumber\\
		&\times \left[ -E_n +\frac{\tr(He^{-p\beta H})}{\tr(e^{-p\beta H})} \right] \nonumber \\
		& -p \frac{e^{-p\beta E_n}}{\tr(e^{-p\beta H})}
		\left[ -E_n + \frac{\tr(He^{-p\beta H})}{\tr(e^{-p\beta H})} \right] \nonumber \\
		&- p \frac{\tr(He^{-p\beta H})}{\tr(e^{-p\beta H})}
		+ p \frac{\tr(e^{-p\beta H})}{\tr(e^{-p\beta H})}\frac{\tr(He^{-p\beta H})}{\tr(e^{-p\beta H})}.
	\end{align}
	
	The last two terms cancel, and we group the remaining terms together and divide by $p$ to obtain
	
	\begin{align}
		0
		&= \left[E_n - \frac{\tr(He^{-p\beta H})}{\tr(e^{-p\beta H})} \right] \\
		&\times \left( 
		\left( \frac{1}{\|\rho\|_p} - \frac{e^{-\beta E_n}}{\|e^{-\beta H}\|_p}\right)^{p-1} \frac{e^{-\beta E_n}}{\|e^{-\beta H}\|_p}
		+ \frac{e^{-p\beta E_n}}{\tr(e^{-p\beta H})}
		\right). \nonumber
	\end{align}
	
	This is zero if and only if
	
	\begin{align}\label{eq:beta_mic}
		E_n = \frac{\tr(He^{-p\beta H})}{\tr(e^{-p\beta H})}.
	\end{align}
	
	By the one-to-one correspondence of energies and canonical inverse temperatures there exists exactly one $\beta$ for a given $E_n$ which obeys \eqref{eq:beta_mic}. This $\beta$ minimizes \eqref{eq:diff_p} and we call it $\beta_E$. It is related to the canonical inverse temperature $\beta_C$, which is defined as the unique solution to \eqref{eq:canoncial_temp},
	via $\beta_C = p \times \beta_E$.
	
	\subsubsection{Value of the minimum} \label{sec:min_val}
	
	To allow for the case of using a microcanonical DM in place of the eigenstate DM (\ref{sec:Microcanonical_DM}), we consider the distance \eqref{eq:diff_p} with
	$\rho$ now of the form \eqref{eq:microcan_state} ($\mathcal{N}=1$ gives eigenstate temperature).
	We assume that $||e^{-\beta H}||_p \geq ||\rho||_p e^{-\beta E_j}$, and we separate the final sum into the difference of two sums.
	
	\begin{align}
		d_p^p(\rho, e^{-\beta H})
		=& \sum_{E_j\in \Delta E}  \left( \frac{1}{\mathcal{N}^{1/p}} - \frac{e^{-\beta E_j}}{\|e^{-\beta H}\|_p} \right)^{p} \nonumber\\&
		+ \sum_{ E_j } \frac{e^{-p\beta E_j}}{\|e^{-\beta H}\|_p^p} - \sum_{ E_j \in \Delta E } \frac{e^{-p\beta E_j}}{\|e^{-\beta H}\|_p^p}.
	\end{align}
	
	Now we consider $\rho$ is constructed from states in the middle of the spectrum, hence we take $\beta$ close to zero, and we can approximate $e^{-\beta E_j} \approx 1$,
	
	\begin{align}
		d_p^p(\rho, e^{-\beta H})
		=& \ \frac{1}{\mathcal{N}} \sum_{E_j\in \Delta E} \left( 1 - ({\mathcal{N}}/{D})^{1/p} \right)^{p} \nonumber\\&
		+ 1 - \sum_{ E_j \in \Delta E } \frac{1}{D} \notag \\
		=& \ \left( 1 - ({\mathcal{N}}/{D})^{1/p} \right)^{p} 
		+ 1 - \frac{\mathcal{N}}{D}. \label{eq:d_p_approx}
	\end{align}
	
	If $p=1$, and we assume $\mathcal{N} \ll D$, it is clear from \eqref{eq:d_p_approx} that $d_1 \approx 2$.
	
	For $p\geq2$ we use the binomial expansion on \eqref{eq:d_p_approx}, and let $D_E = \mathcal{N}/D$,
	\begin{align}
		\left( 1 - D_E^{1/p} \right)^{p} 
		= \sum_{n=0}^{\infty} \binom{p}{n} (-1)^n D_E^{n/p} 
	\end{align}
	Resulting in 
	\begin{align}
		d_p^p(\rho, e^{-\beta H})
		= 2 - pD_{E}^{1/p} + O\left(D_{E}^{\ell}\right)
	\end{align}
	
	Here, $\ell = \min(1,2/p)$. Then finally to obtain $d_p$, we raise both sides to $1/p$, and use the binomial expansion again,
	\begin{align}
		d_p(\rho, e^{-\beta H})
		&= 2^{1/p} - 2^{1/p-1}D_{E}^{1/p} + O\left(D_{E}^{\ell}\right)
	\end{align}
	
	Thus the leading perturbation is $D_{E}^{1/p} = (\mathcal{N}/D)^{1/p}$. So when $\mathcal{N}\ll D$, $d_p$ is close to $2^{1/p}$ for bulk eigenstates.

	
	\section{Subsystem Temperature}\label{sec:Subsystem_Temperature}
	
	We  now turn to  the subsystem temperature, which we have defined as
	\begin{equation}
		\beta_{S} = \displaystyle\argmin_{\beta} d_p\left(\rho^A,\rho_C^A\right).
	\end{equation} 
	Here, $\rho^{A}= \tr_{B}(\rho)$, with $\rho$ an eigenstate DM, and $\rho_{C}^{A}= \tr_{B}(\rho_{C})$.  The partial trace prevents a calculation
	similar to that we used to derive $\beta_{E}$ = $\beta_{C}/p$; we thus do not have analytical
	predictions for the relationship between $\beta_{S}$ and $\beta_{C}$.  On physical grounds, one
	expects $\beta_{S}$ to match $\beta_{C}$ for $L_A\ll{L}$ and large $L$.
	We first present our numerical findings for $\beta_{S}$ in various quantum systems, exploring this expected correspondence. Following this, we present an analytical argument for how the distance $d_p\left(\rho^A,\rho_C^A\right)$, at infinite temperature, should scale in the limit of $L_A\ll{L}$ and large $L$.
	
	\subsection{Main Results}
	
\begin{figure}
		\includegraphics[width=\linewidth]{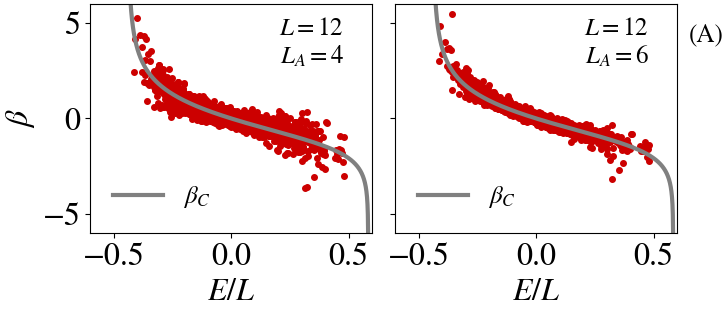}
		\includegraphics[width=\linewidth]{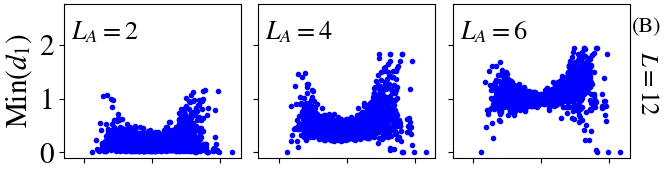}
		\includegraphics[width=\linewidth]{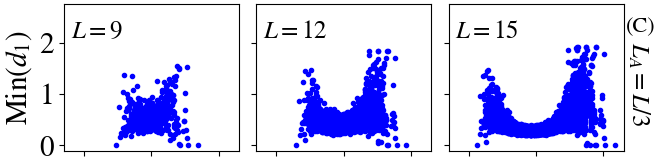}
		\includegraphics[width=\linewidth]{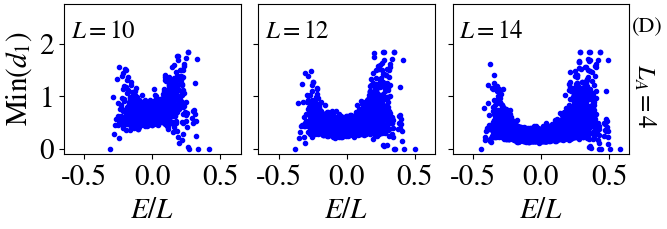}
		\caption{Subsystem temperature results for the chaotic Ising model with $h_z = 0.5$ and $h_x = 0.75$.
			\textbf{(A)} $\beta$ that minimizes $d_1(\rho^{A},\rho_{C}^{A})$ ($\beta_{S}$) versus energy, plotted along side the canonical $\beta_{C}$ curve, for the given $L$ and $L_A$.
			\textbf{(B)-(D)} $\min(d_1(\rho^A,\rho^A_C))$ plotted versus energy, each row illustrating a different scaling of system/subsystem size.}
		\label{fig:SubT_EvsB_1}
	\end{figure}

	The values of $\beta_{S}$ are found in general to be scattered around $\beta_{C}$, as shown in
	Fig.~\ref{fig:SubT_EvsB_1}(A) for the chaotic Ising model.  The width of this scatter generally
	decreases with system size (both $L_A$ and $L$), as quantified further below.
	In stark contrast to $\beta_E$, there is no obvious dependence on the distance measure used --- the
	qualitative behavior is the same for all $p$ except $p=\infty$, see Appendix \ref{sec:altp} for $p=2$ data. We therefore present
	numerical results for the trace distance, $p=1$.

\begin{figure}
		\includegraphics[width=\linewidth]{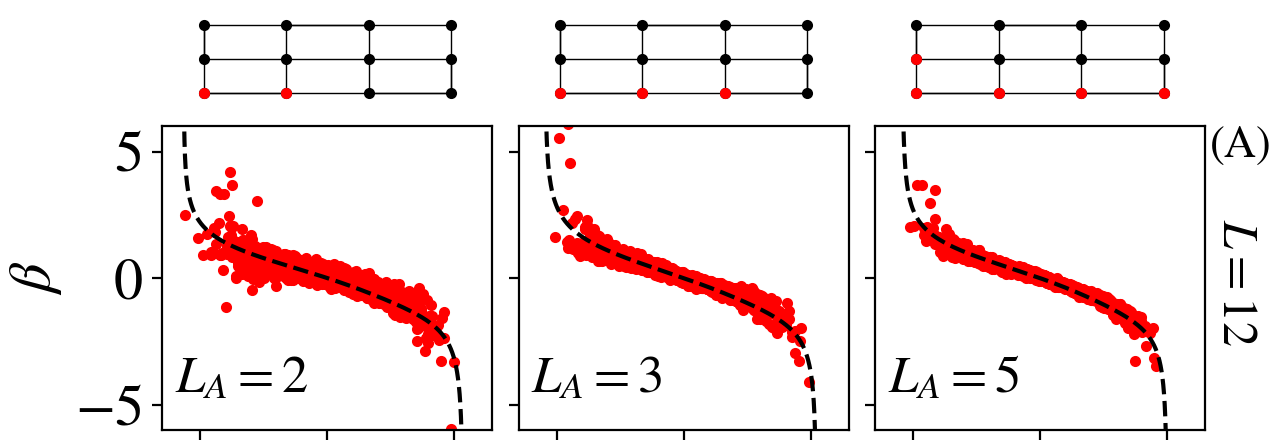}
		\includegraphics[width=\linewidth]{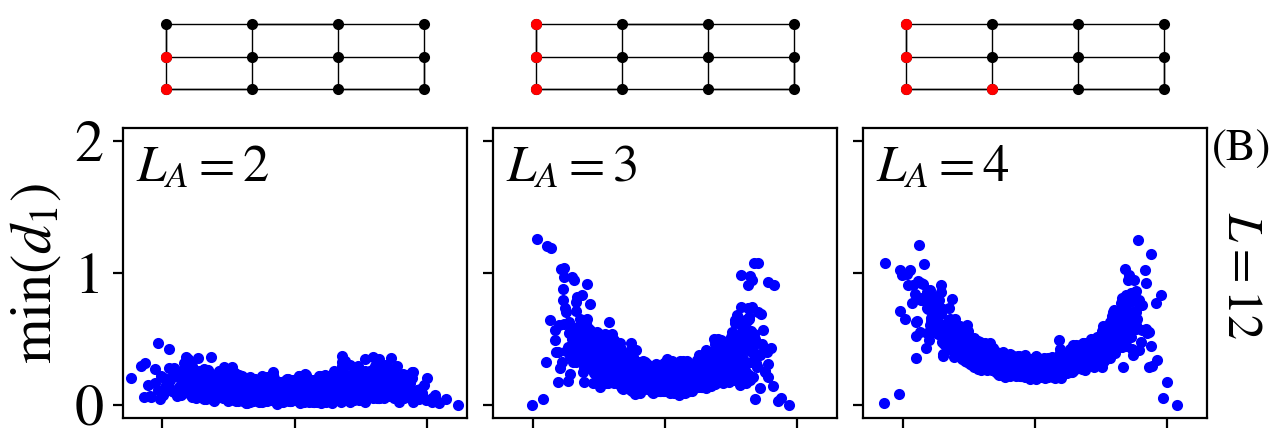}
		\includegraphics[width=\linewidth]{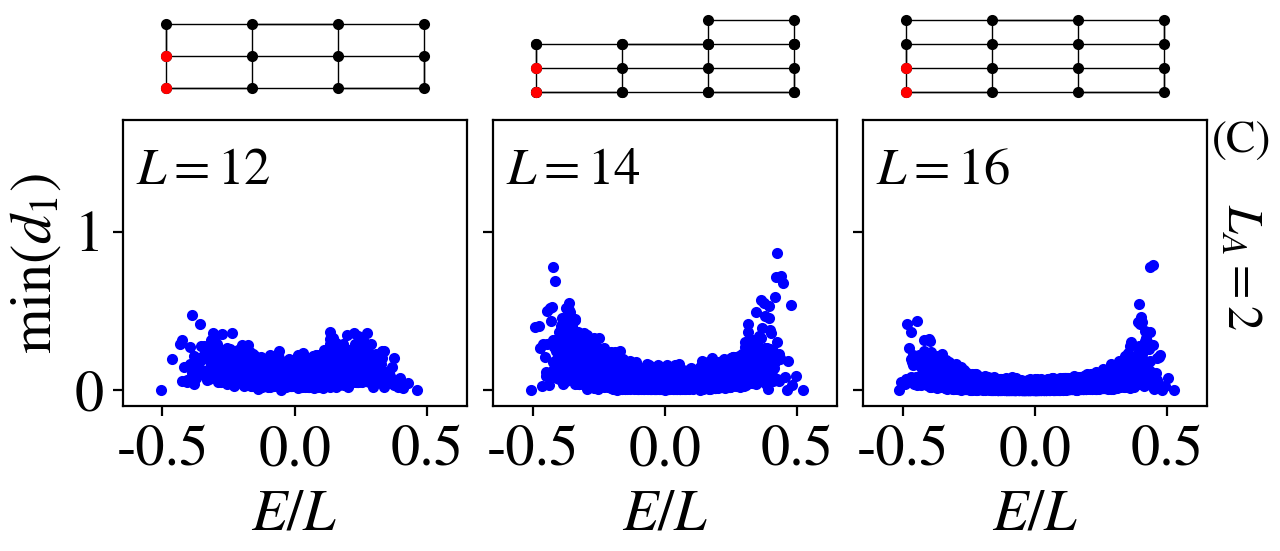}
		\caption{Subsystem temperature results for square lattice model with the addition of staggered $S^x$ fields with $h_x = 0.5$. The geometry of each system is illustrated above each plot, in which, red and black sites correspond to the subsystems $A$ and $B$ respectively. The results in each plot are of a particular realization - \textbf{(A):} $\beta$ that minimizes $d_1(\rho^{A},\rho_{C}^{A})$ ($\beta_{S}$) versus energy, plotted alongside the canonical $\beta_{C}$ curve, for the given $L$ and $L_A$.  \textbf{(B-C):} $\min(d_1(\rho^{A},\rho_{C}^{A}))$ plotted versus energy, each row illustrating a different scaling of system/subsystem size. }
		\label{fig:eth_2D}
	\end{figure}

	The qualitative results of Fig.~\ref{fig:SubT_EvsB_1} are not specific to 1D chains. This is clear from the strikingly similar results we obtain for the 2D square lattice model as shown in Fig.~\ref{fig:eth_2D}.
	In Fig.~\ref{fig:eth_2D}, we illustrate the geometry of the square lattice for each given system/subsystem parameters, alongside the respective $\beta_{S}$ and $\min(d_1)$ versus $E$ plots. In the geometry illustrations, the red and black points represent the subsystems $A$ and $B$ respectively. We observe similar results to that of a chaotic 1D spin chain such as those in Fig.~\ref{fig:SubT_EvsB_1}.
	
	When increasing $L_A$ with fixed total system size
	$L$, the variance of $\beta_{S}$ and the distance between $\beta_{S}$
	and $\beta_{C}$ decrease, up to $L_A=L/2$.  For $L_A>L/2$ the
	distribution of $\beta_{S}$ values changes shape and shows additional
	features, perhaps resulting from $\rho^A$ no longer having full rank.
	See Appendix \ref{sec:subsys_data} for examples of results from systems with $L_A > L/2$.
	
	Although $|\beta_{S}-\beta_{C}|$ and the variance of $\beta_{S}$ improve with increasing $L_A$, the
	minimum distance between $\rho^A$ and $\rho_C^A$ does not, as is visible from Figures \ref{fig:SubT_EvsB_1}(B) and \ref{fig:eth_2D}(B).  The
	average $\min(d_1)$ increases markedly with $L_A$.  The reduced DM has decreasing resemblance to the
	reduced canonical DM, presumably because of the decreasing size of the complement $B$, which plays
	the role of a bath.
	
	Increasing $L$ while keeping the fraction $L_A/L$ fixed, we again find the variance  of $\beta_{S}$
	to decrease.  In this limit, $\min(d_1)$ on average decreases when the fraction  $L_A/L$ is
	$<\frac{1}{2}$ (see Fig.~\ref{fig:SubT_EvsB_1}(C)), and is remarkably stable as a function of $L$ when the
	fraction is $L_A/L= \frac{1}{2}$, see Appendix \ref{sec:subsys_data}.
	
\begin{figure}
		\includegraphics[width=\linewidth]{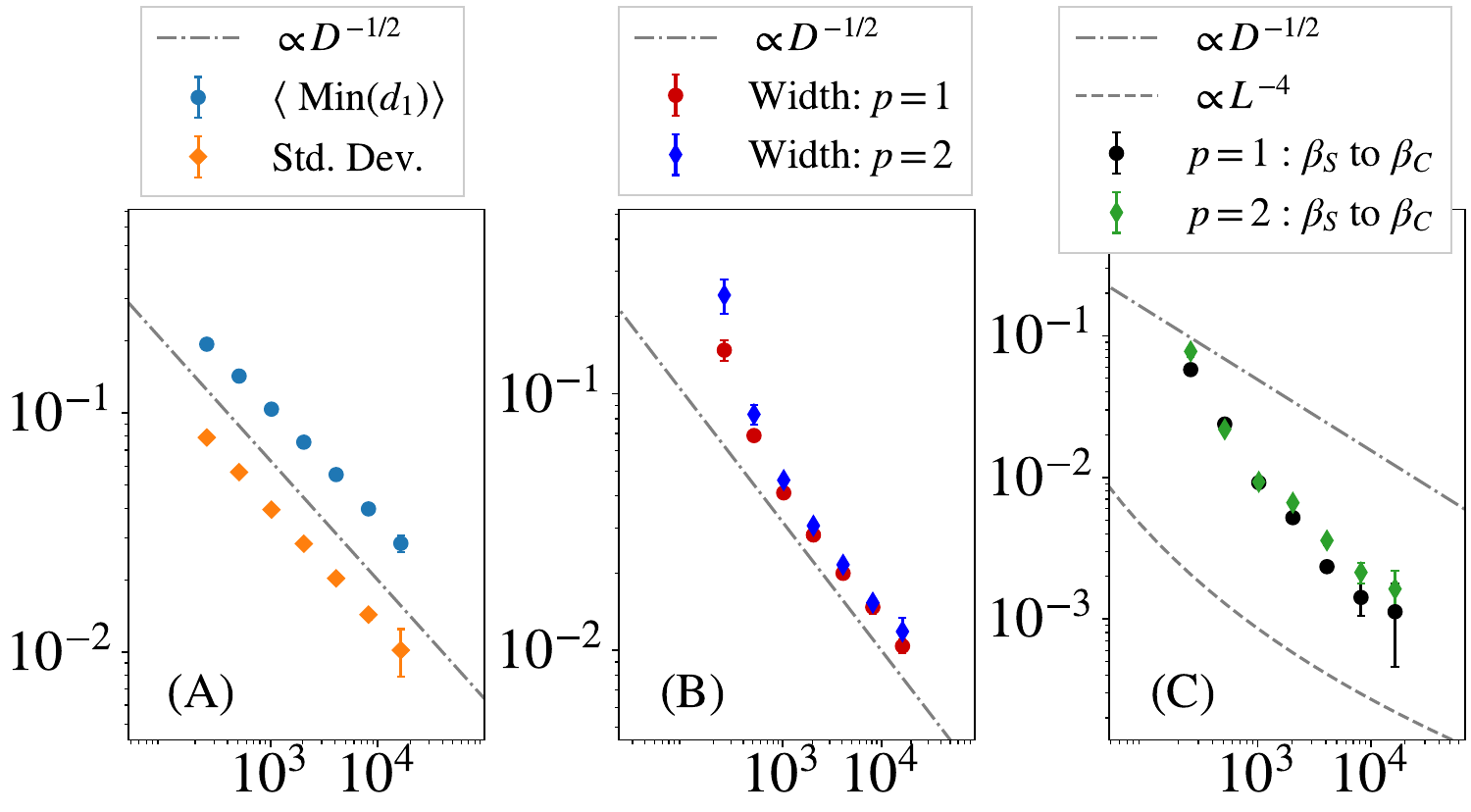}
		\includegraphics[width=\linewidth]{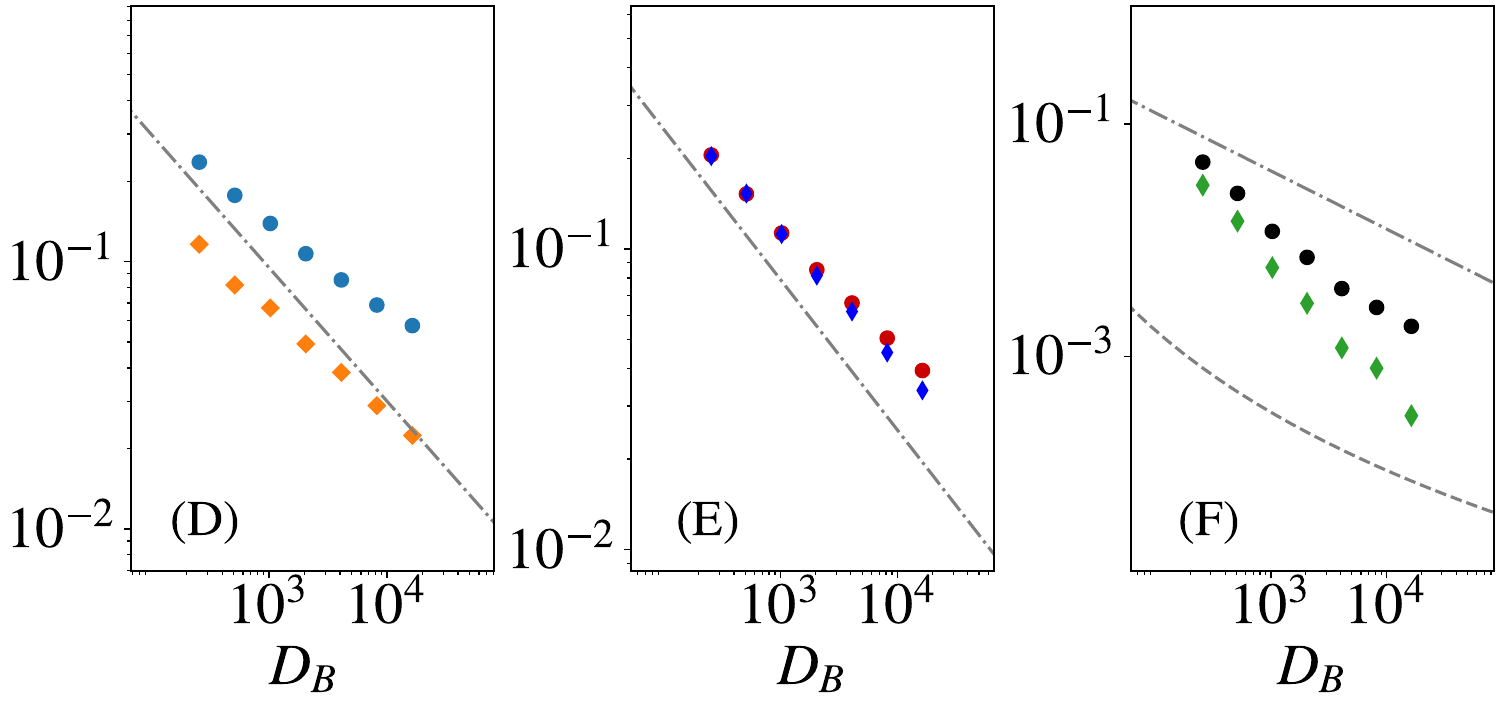}
		\caption{Subsystem temperature scaling for \textbf{(A)-(C)} disordered field \textit{XXZ}-chain with $W = 0.25$, and $L_A=2$, over many disorder realizations.
			\textbf{(D)-(F)} chaotic Ising model with $h_x = 0.75$, $h_z = 0.5$, and $L_A=2$. 
			For both models, statistics are taken from the central 20\% of the spectrum.
			\textbf{(A),(D)} Mean of $\min(d_1(\rho^A,\rho^A_C))$ and its
			standard deviation, vs. $D_B$ for $p=1$.  
			\textbf{(B),(E)} Width of $\beta_{S}$ vs. $D_B$ for $p=1,2$.  
			\textbf{(C),(F)} RMS-distance from the linear fit of $\beta_{S}$, to
			$\beta_{C}$ curve versus $D_B$, for $p=1,2$.
		}
		\label{fig:SubT_scaling}
	\end{figure}

	We now consider fixed $L_A$ and increasing $L$ (or increasing $L_B=L-L_A$).  The reduced DMs become
	increasingly similar in this limit, as shown in Figures \ref{fig:SubT_EvsB_1}(D) and \ref{fig:eth_2D}(C).  In Fig.~\ref{fig:SubT_scaling} we
	show scaling behaviors in this limit computed using the central 20\% of the spectrum.
	Fig.~\ref{fig:SubT_scaling} (A)-(C) shows results for the disordered-field \textit{XXZ}-chain, while Fig.~\ref{fig:SubT_scaling} (D)-(F) shows those for the chaotic Ising model, both with $L_A=2$.
	
	The minimum distance between DMs $\rho_{C}^A$ and $\rho^A$ decreases apparently exponentially with
	system size, consistent with the upper bound ${\sim}D_B^{-1/2}$ (equivalently ${\sim}D^{-1/2}$), see Fig.~\ref{fig:SubT_scaling}(A)/(D).  While this scaling is difficult to prove for a general Hamiltonian,
	one can argue for this dependence based on assuming the eigenstates to be effectively random
	Gaussian states near the center of the spectrum.  
	This is known to be a good but not
	perfect approximation for chaotic many-body systems with local interactions
	\cite{Beugeling_Haque_PRE2018, Khaymovich_Haque_McClarty_PRL2019, Baecker_Haque_Khaymovich_PRE2019,
		DeTomasi_Khaymovich_Pollmann_Warzel_PRB2021, Prosen_Sotiriadis_PRL2021,
		Ueda_LocalRMT_PRL2021,Buchleitner_BH_spectrum_PRL2021}, and has been used to analyze ETH
	\cite{Srednicki_PhysRevE.50.888, Neunhahn_Marquardt_PRE2012, BrandinoKonikMussardo_PRB12,
		Beugeling_Moessner_Haque_PRE2014, Beugeling_Moessner_Haque_PRE2015, Alessio_Rigol_AdvPhys2016,
		Mondaini_Rigol_2DIsing_PRE2017, Khaymovich_Haque_McClarty_PRL2019, Mierzejewski_Vidmar_PRL2020,
		Nakerst_Haque_PRE2021}.  With this assumption, the reduced DM is a Wishart matrix, while
	the infinite-temperature canonical DM is an identity matrix.  Thus the question is, how fast a
	$p$-normalized Wishart matrix concentrates around an identity matrix?  Using concentration of measure
	results \cite{book_Vershynin_highDprob_2018}, one can show that this dependence is at most
	$D_{B}^{-1/2}$, as shown in the following Section \ref{sec:subscaling}.  
	
	The width of $\min(d_1)$ clouds appears to decrease at least as fast as $\sim D_B^{-1/2}$ as well,
	as shown in Fig.~\ref{fig:SubT_scaling}(A)/(D).  This is reasonable as $d_1$ is bounded from below and the average
	$\min(d_1)$ decreases as $\sim D_B^{-1/2}$.
	
	The width of the $\beta_{S}$ values which minimize $d_1$ also appears to have $\sim D_B^{-1/2}$
	scaling (at most), see Fig.~\ref{fig:SubT_scaling}(B)/(E).  We have been unable to formulate an analytic argument for this scaling. 
	As the width of the $\beta_{S}$ cloud decreases, these values concentrate on a line in the
	$L\to\infty$ limit.  Fig.~\ref{fig:SubT_scaling}(C)/(F) shows, by plotting the average distance of the
	$\beta_{S}$ cloud to the $\beta_C$ line, that the asymptotic shape of the $\beta_{S}$ cloud
	coincides with the $\beta_C$ line.  From the available data, it is unclear whether this approach is
	power-law or exponential in $L$.  Again, no analytical prediction is currently available for this
	dependence.  In Ref.~\cite{Dymarsky_Lashkari_Liu_PRE2018}, an upper-bound scaling of $L^{-1}$ is
	derived for a closely related quantity, namely, $d_1(\rho^{A},\rho_{C}^{A})$ evaluated at
	$\beta_{C}$, instead of at its minimum $\beta_{S}$.  Fig.~\ref{fig:SubT_scaling}(C)/(F) shows that the actual scaling
	of $\min{d_1}$ is much faster. In Appendix \ref{sec:betaC} we calculate the average value of $d_1(\rho^{A},\rho_{C}^{A})$ at $\beta_{C}$ as a function of $L$.
	
	
	\subsection{Scaling of subsystem distance derivation} \label{sec:subscaling}
	In this subsection we will prove that at infinite temperature the Schatten-1 distance between an eigenstate of a generic Hamiltonian and the reduced canonical density matrix decreases as $O(D^{-1/2})$ or equivalently $O(D_B^{-1/2})$ in the limit of fixed subsystem size $L_A$ and increasing complement $L_B\to\infty$. Recall that $D_B = 2^{L_B}$ and $D = 2^L$, so fixed $L_A$ and increasing $L_B$ (increasing $L$) is equivalent to fixed $D_A$ and increasing $D_B$ (increasing $D=D_AD_B$).
	
	At infinite temperature the canonical density matrix $\rho_C$ is the maximally entangled state $\rho_C = D^{-1}\cdot \mathbb{1}_{D}$ and its reduced density matrix is $\rho_C^A = D_A^{-1}\cdot \mathbb{1}_{D_A}$. So the spectrum of $\rho^A-\rho_C^A$ equals the spectrum of $\rho^A$ shifted by the constant $D_{A}^{-1}$. The Schatten-1 distance between the reduced eigenstate density matrix and the reduced canonical density matrix can then be written as
	\begin{equation}
		\left|\left|\rho^A-\frac{\tr_{B}e^{\beta H}}{\tr({e^{-\beta H}})}\right|\right|_1 
		=\ \sum_{j=1}^{D_A} \left|\lambda_j - \frac{1}{D_A}\right|,
		\label{eq:est_tr_diff}
	\end{equation}
	where the $\lambda_j$ denote the eigenvalues of $\rho^A$.
	
	We assume that an eigenstate $\ket{E}$ of a generic Hamiltonian at infinite temperature is well approximated by a random state uniformly distributed on the $S^{D-1}$ sphere. For large $D$ the uniform distribution on $S^{D-1}$ is close to a multivariate Gaussian distribution with independent components and mean 0 and variance $D^{-1}$. Because the density matrix $\outerprod{E}{E}$ has rank 1 the reduced density matrix $\rho^A$ is given by $\rho^A = D^{-1}XX^t$, where $X$ is a $D_A\times D_B$ matrix with independent Gaussian entries with mean 0 and variance 1. The reduced eigenstate density $\rho^A$ is distributed according to the Wishart distribution with expectation value $D_A^{-1}\cdot \mathbb{1}_{D_A} = \rho_C$. So the problem of finding an upper bound for 
	\eqref{eq:est_tr_diff} reduces to finding an upper bound on how quickly Wishart matrices concentrate around their mean.
	
	To answer this question we use a concentration of measure result about singular values of
	Gaussian rectangular matrices $X$, which can be found in, e.g., \cite{book_Vershynin_highDprob_2018} (Corollary 7.3.3 and exercise 7.3.4).
	For $0<t$ with probability $1-2e^{-t^2/2}$ all singular values $\sigma_j$ of $X$ obey
	\begin{equation}
		\sqrt{D_B} - \sqrt{D_A} - t \le \sigma_j \le \sqrt{D_B} + \sqrt{D_A} + t.
	\end{equation}
	The eigenvalues $\lambda_j$ of $\rho^A = D^{-1}XX^t$ are the squared singular values of $X$, re-normalized by $D^{-1}$, namely $\lambda_j = D^{-1} \sigma_j^2$. So for $0<s<1 +  D_A/D_B - 2 \sqrt{D_A/D_B}$ with probability $1-2e^{-sD_B/2}$ we have 
	\begin{equation}
		\sum_{j=1}^{D_A} \left|\lambda_j - \frac{1}{D_A}\right| \le \frac{D_A}{D_B} + 2\frac{\sqrt{D_A}}{\sqrt{D_B}} + s + 2\left( 1 + \frac{\sqrt{D_A}}{\sqrt{D_B}}\right) \sqrt{s}.
	\end{equation}
	Note that for fixed $D_A$ the leading order in the $s$ independent term is $D_B^{-1/2}$. Under some mild assumptions on higher moments of $\lambda_j$, for example that the second moment of $\lambda_j$ increases at most polynomially for fixed $D_A$ and increasing $D_B$, we can asymptotically estimate the expected value of 
	\eqref{eq:est_tr_diff} as
	\begin{equation}
		E\left[ \sum_{j=1}^{D_A} \left|\lambda_j - \frac{1}{D_A}\right| \right]
		\lessapprox 2(\sqrt{D_A} + 1) D_B^{-1/2} + O(D_B^{-1}).
	\end{equation}
	Thus one expects the Schatten-1 distance between the reduced density matrix of a Gaussian random state and the maximally mixed state to decrease as $O(D_B^{-1/2})$ or equivalently $O(D^{-1/2})$ for fixed $D_A$ and increasing $D_B$.

	
	\section{Alternate formulations}\label{sec:Alternate_formulations}
	Here, we present some possible alternate formulations of our eigenstate-based temperatures.
	First, we discuss using the Bures distance in place of the Schatten $p$-distance. We derive an analytical result for the eigenstate temperature utilizing the Bures distance, analogous to that shown in \ref{sec:eigenstate_temperature}. Following this, we discuss the use of $\exp(-\beta H_A)$ in place of $\tr(\rho_{C})$ in the subsystem temperature. We provide numerical results for this alternate formulation of $\beta_{S}$.
	
	
	\subsection{Bures Distance}\label{sec:FidBure}
	Instead of the Schatten $p$-distances, one could justifiably use the Bures
	distance, related to the fidelity \cite{book_Barnett_QuantInfo_2009, nielsen_chuang_2010}. 
	We have found that the subsystem temperature $\beta_S$, when calculated using the Bures distance, has the same overall features as found using the Schatten distances. 
	
	Additionally, the eigenstate temperature if based on the Bures distance, is the
	same as $\beta_C$, i.e., the same as $\beta_E$ for $p=1$. We derive this analytically below, and also illustrate the result numerically.  
	
	The fidelity between two density matrices is given as
	\begin{equation}\label{eq:fidelity_def}
		F(\rho,\sigma) = (\tr\sqrt{\rho^{1/2}\sigma\rho^{1/2}})^2,
	\end{equation}
	or sometimes as the square root fidelity (quantity fidelity) $F'(\rho,\sigma) = \sqrt{F(\rho,\sigma)}$.
	It is a measure of how similar $\rho$ and $\sigma$ are, but it is not a metric on density operators. It is symmetric in the inputs, and is bounded between 0 and 1.
	
	Before delving into maximizing $F$, we note that the square root of a microcanonical density matrix $\rho$, as defined in \eqref{eq:microcan_state}, is $\sqrt{\mathcal{N}}\rho$, as 
	\begin{align}
		(\sqrt{\mathcal{N}}\rho)^2 &= \frac{\mathcal{N}}{\mathcal{N}^2} \sum_{E_j,E_{j'} \in \Delta E} \ket{E_j}\innerprod{E_j}{E_{j'}}\bra{E_{j'}} \\
		&= \frac{1}{\mathcal{N}} \sum_{E_j \in \Delta E} \outerprod{E_j}{E_j} = \rho
	\end{align}
	Now, we want to maximize the fidelity between a microcanonical state $\rho = \rho_{MC}$ and a canonical state $\rho_{C}.$
	\begin{align}
		F(\rho,\rho_{C}) &= \tr(\sqrt{\rho^{1/2}\rho_{C}\rho^{1/2}})^2 = (\tr\sqrt{\rho\rho_{C}})^2\\ 
		&= \left(\tr(\sqrt{\mathcal{N}}\rho e^{-\beta H/2})\right)^2/\tr(e^{-\beta H})
	\end{align}
	\begin{align}
		&= \frac{1}{\mathcal{N}\tr(e^{-\beta H})}\left(\tr( \sum_{E_{j} \in \Delta E} e^{-\beta E_{j}/2}\outerprod{E_{j}}{E_{j}} )\right)^2\\
		&= \frac{1}{\mathcal{N}\tr(e^{-\beta H})}\left( \sum_{E_{j},E_{j'} \in \Delta E} e^{-\frac{\beta}{2} (E_{j}+E_{j'})} \right)
	\end{align}
	Now to find the value of $\beta$ which maximizes $F(\rho,\rho_{C})$, we simply differentiate to obtain
	\begin{align}
		\frac{\partial F}{\partial \beta}& = \frac{\tr(He^{-\beta H})}{\mathcal{N}\tr(e^{-\beta H})^2}\sum_{E_j,E_{j'} \in \Delta E}e^{-\frac{\beta}{2}(E_j+E_{j'})}\notag\\
		&+ \frac{1}{\mathcal{N}\tr(e^{-\beta H})}\sum_{E_{j},E_{j'} \in \Delta E}-\frac{(E_j+E_{j'})}{2}e^{-\frac{\beta}{2}(E_j+E_{j'})}
	\end{align}
	\begin{align}
		\frac{\partial F}{\partial \beta} = &\frac{1}{\mathcal{N}\tr(e^{-\beta H})^2} \sum_{E_{j},E_{j'} \in \Delta E}e^{-\frac{\beta}{2}(E_j+E_{j'})}\times\notag\\
		&\times\left( \tr(H e^{-\beta H}) - \frac{E_j+E_{j'}}{2}\tr(e^{-\beta H} )\right)
	\end{align}
	We then make the approximation of $E_j \approx E_{j'} \approx E$ for $E_j,E_{j'} \in \Delta E$, which is accurate for small $\Delta E$, and is exact when $\Delta E$ contains a single eigenstate.
	\begin{align}
		\frac{\partial F}{\partial \beta} = \frac{e^{-\beta E}}{\tr(e^{-\beta H})}\left( \frac{\tr(H e^{-\beta H})}{\tr(e^{-\beta H})} - E \right)
	\end{align}
	Then setting this equal to zero, we find the only roots of the equation are when
	\begin{align}
		E = \frac{\tr(H e^{-\beta H})}{\tr(e^{-\beta H})}.
	\end{align}
	This is the canonical energy-temperature relation \eqref{eq:canoncial_temp}, meaning that the temperature which maximizes the fidelity between a microcanonical state $\rho$ with energy $E$, and a canonical state, is in fact the canonical temperature $\beta_{C}$.
	
	The Bures distance is defined as
	\begin{equation}
		d_B(\rho,\sigma)^2 = 2(1-\sqrt{F(\rho,\sigma)}),
	\end{equation}
	with $F(\rho,\sigma)$ defined as \eqref{eq:fidelity_def}. The Bures distance is minimized when the Fidelity is maximized (i.e. when $F = 1$). Thus the Bures distance is minimized when $\beta = \beta_{C}$ also.
	
\begin{figure}
		\includegraphics[width=\linewidth]{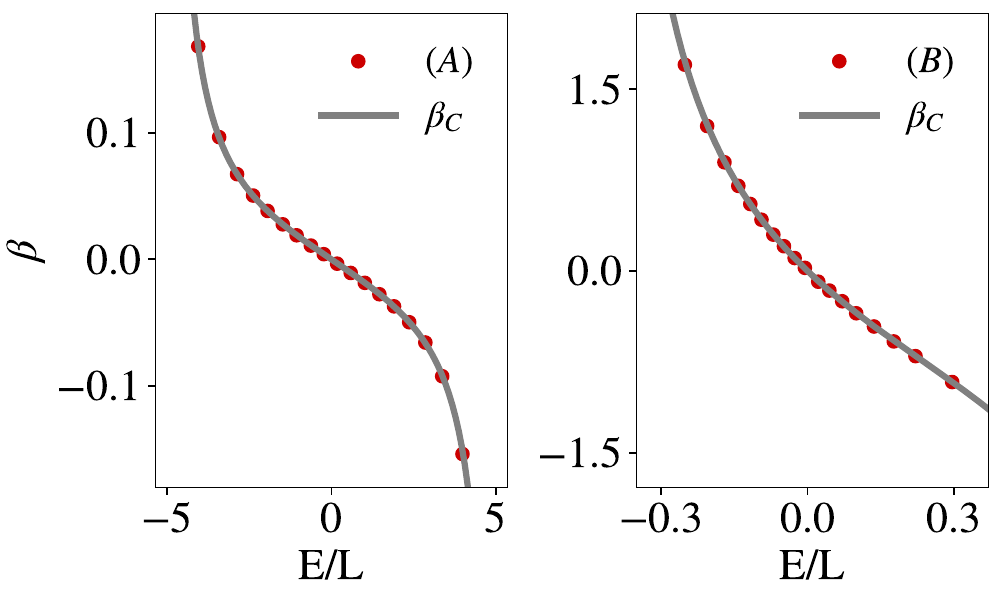}
		\caption{Finite window eigenstate temperature $\beta_{\Delta{E}}$ calculated using Bures distance $d_B(\rho_{MC},\rho_C)$, for two models: $\mathbf{(A)}$ Random, real and symmetric matrix, $\mathbf{(B)}$ Chaotic Ising model with $h_x = 0.5$, $h_z = 0.75$. In both cases, $L = 9$ ($D=2^9$) and 20 energy windows are uniformly chosen from the spectrum of the given Hamiltonian.}
		\label{fig:bures}
	\end{figure}

	We numerically demonstrate this result in Fig.~\ref{fig:bures}. We present results for both the chaotic Ising model used previously, and also for a random real symmetric matrix, both clearly illustrating the model independent result $\beta_{\Delta{E}} = \beta_{C}$ for the Bures distance $d_B(\rho_{MC},\rho_{C})$.
	
	\subsection{Local Hamiltonian density matrix}
	For the subsystem temperature, we compared $\rho^{A}$ to $\rho_{C}^{A}=\tr_{B}\exp(-\beta H)$.  An
	obvious alternative is to compare with $\exp(-\beta{}H_{A})$.  If $H_{AB}$ is nonzero, the two
	are not equivalent, as discussed widely in the literature \cite{
		Hilt_HamiltonianOfMeanForce_PRE2011,Jarzynski_JSTAT2004,
		Dong_Liu_Sun_couplings_PRA2007, Gelin_Thoss_subensemble_PRE2009,
		Riera_Gogolin_Eisert_PRL2012, Sun_systembathcoupling_PRE2014,
		Nazir_QuantumHeatEngine_PRE2017, Dymarsky_Lashkari_Liu_PRE2018,
		PerarnauLlobet_Riera_Eisert_StrongCoupling_PRL2018, Campisi_Talkner_Haenggi_PRL2009,
		Seifert_strongcoupling_PRL2016, Strasberg_2016, Philbin_2016}, 
	e.g., in the context of extracting
	an effective ``Hamiltonian of mean force'' for the subsystem
	\cite{Hilt_HamiltonianOfMeanForce_PRE2011,Kirkwood_1935,Jarzynski_JSTAT2004, 		Campisi_Talkner_Haenggi_PRL2009, Seifert_strongcoupling_PRL2016, 
		Strasberg_2016,Philbin_2016}.  Numerically, we have found that using $\exp(-\beta H_A)$ to define $\beta_S$ leads
	to very similar results to those obtained using $\rho_{C}^{A}$, except for eigenstates at the
	spectral edges. 
	
\begin{figure}
		\includegraphics[width=\linewidth]{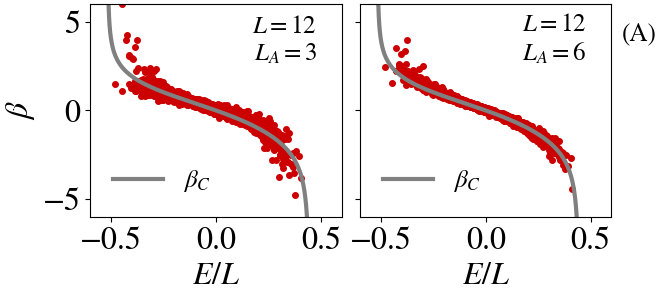}
		\includegraphics[width=\linewidth]{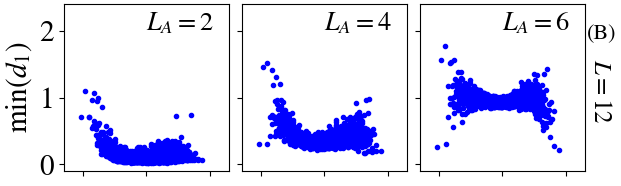}
		\includegraphics[width=\linewidth]{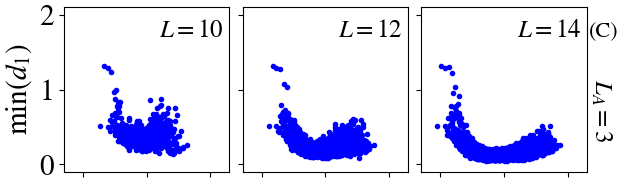}
		\includegraphics[width=\linewidth]{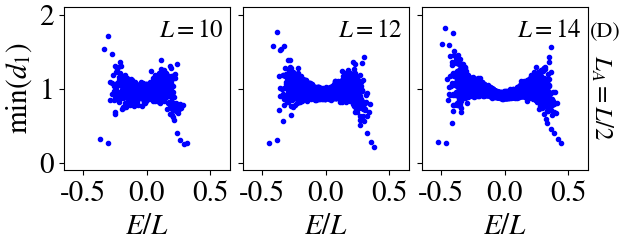}
		\caption{Subsystem temperature results with $\rho^{A} = \exp(-\beta H_A)$, for staggered field model with $h_x = h_z = 0.5$, $J=1$ and $\Delta=0.95$.
			\textbf{(A)} $\beta$ minimizing $d_1(\rho^{A},\rho_{C}^{A})$ ($\beta_{S}$) versus energy, plotted along side the canonical $\beta_{C}$ curve, for the given system/subsystem size.
			\textbf{(B)-(D)} $\min(d_1(\rho^A,\rho^A_C))$ plotted versus energy, each row illustrating a different scaling of system/subsystem size.}
		\label{fig:localH_EvsB}
	\end{figure}

\begin{figure}
		\includegraphics[width=\linewidth]{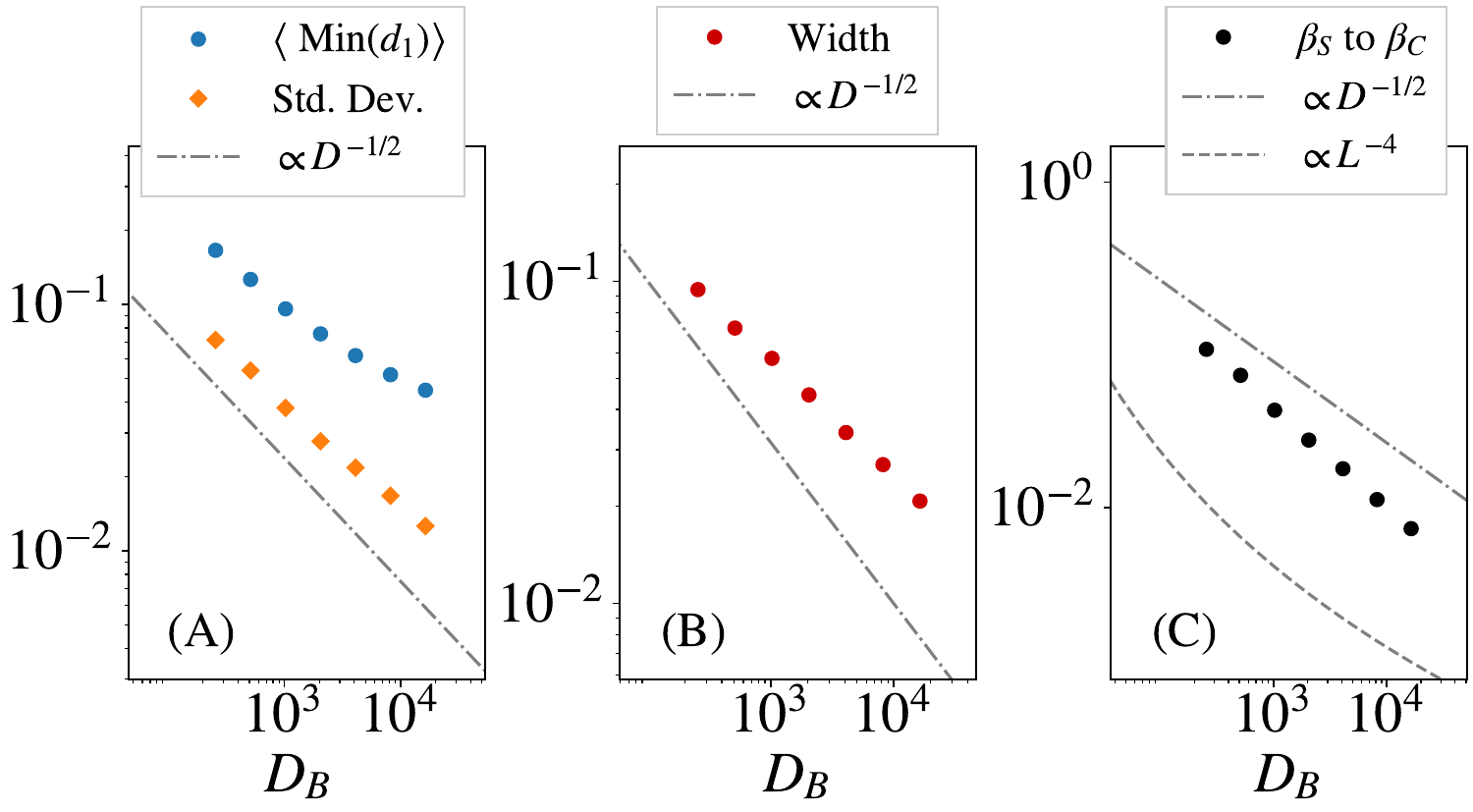}
		\caption{Subsystem temperature results with $\rho^{A} = \exp(-\beta H_A)$, for staggered field \textit{XXZ}-chain with $J=1$, $\Delta=0.95$, $h_x = h_z = 0.5$ and $L_A = 2$. Statistics from the central 20\% of the spectrum. 
			\textbf{(A)} Mean of $\min(d_1(\rho^A,\rho^A_C))$ and its
			standard deviation, vs. $D_B$ for $p=1$.
			\textbf{(B)} Width of $\beta_{S}$ vs. $D_B$ for $p=1,2$.
			\textbf{(C)} RMS-distance from the linear fit of $\beta_{S}$, to
			$\beta_{C}$ curve versus $D_B$, for $p=1,2$.}
		\label{fig:localH_scaling}
	\end{figure}

	In Fig.~\ref{fig:localH_EvsB} we illustrate the behavior of $\beta_{S}$ and $\min(d_p(\rho^{A},\rho_{C}^A))$ with $\rho_{C}^{A} = \exp(-\beta H_{A})$. We see the general behavior is the same as in Figures \ref{fig:SubT_EvsB_1} and \ref{fig:eth_2D}.
	In Fig.~\ref{fig:localH_scaling} we also illustrate similar scalings as seen in Fig.~\ref{fig:SubT_scaling}.
	
	\section{Deviation in non-thermalizing systems}\label{sec:non_chaotic}
	
	Up to now, we have been solely concerned with chaotic systems that are expected to thermalize and hence satisfy the ETH (ergodic). The subsystem temperature is based on ETH predictions for density matrices restricted to a local subsystem. One could then ask what happens to the temperature in a system that is expected to violate the ETH, i.e., one which does not thermalize (non-ergodic).
	
\begin{figure}
		\includegraphics[width=0.99\linewidth]{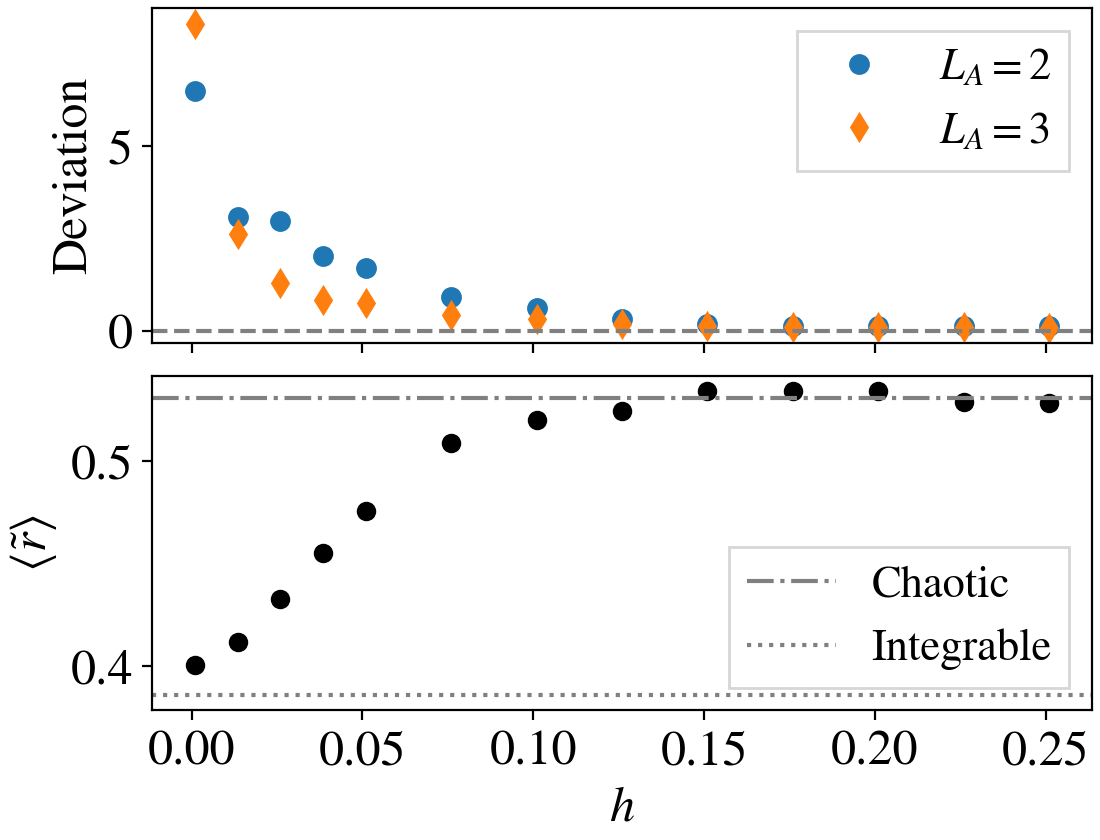}
		\caption{Subsystem temperature results for staggered field model with $L=12$, $J=1$, $\Delta = 0.95$, and $h_z = h_x = h$. 
			\textbf{Top:} The RMS-distance between $\beta_{S}$ and $\beta_{C}$ (Deviation) plotted versus shared field strength $h$. The RMS-distance is calculated for eigenstates in the central 20\% of the spectrum.
			\textbf{Bottom:} Average restricted gap ratio value plotted versus shared field strength $h$.
		}
		\label{fig:non_thermalizing}
	\end{figure}

	In order to investigate this effect, we shall consider the staggered field model with varying field strength $h = h_z = h_x$. For finite, non-zero $h$, the system should in general be thermalizing. Of course, when $h=0$ the system is simply the \textit{XXZ} chain and is known to be exactly solvable via the Bethe ansatz. Thus if we tune $h$, from some finite non-zero value, towards zero, the system should approach a non-thermalizing regime.
	In the top panel of Fig.~\ref{fig:non_thermalizing} we plot the RMS-distance between $\beta_{C}$ and $\beta_{S}$ for such a system as a function of magnetic field strength $h$. As one could expect, when $h\to0$ the deviation between the temperatures increases, due to the system no longer thermalizing. 
	
	To illustrate the systems approach to a non-thermalizing regime, we have plotted the average restricted gap ratio $\langle\tilde{r}\rangle$ against the field strength $h$ in the bottom panel of Fig.~\ref{fig:non_thermalizing}. The restricted gap ratio is defined as the minimum of the gap ratio $r$ and its inverse $r^{-1}$. The gap ratio itself is defined as the ratio of two consecutive level-spacings. Level-spacing statistics are an effective tool in classifying a system as ergodic (chaotic) or non-ergodic (integrable). The restricted gap ratio $\langle\tilde{r}\rangle$ is particularly useful, as it avoids the need to perform an unfolding procedure on the spectrum, as is often required for bare consecutive level-spacings.
	In the bottom panel of Fig.~\ref{fig:non_thermalizing}, we have marked the predicted average restricted gap ratio values for chaotic and integrable systems, $0.5307$ and $0.386$ respectively \cite{AtasBogomolny_RandomMatrixStats_PRL2012}. As expected, $\langle\tilde{r}\rangle$ approaches the predicted value for non-ergodic systems as $h\to{0}$, coinciding with the increasing deviation between $\beta_{S}$ and $\beta_{C}$.
	
	\section{Summary \& Discussion}\label{sec:Discussions}
	
	Our first eigenstate-based temperature, $\beta_E$, turned out to be
	determined solely by the eigenvalues.  It has interesting (arguably unexpected) dependencies on the
	distance measure.  The relation $\beta_E=\beta_C/p$ is a mathematical result that holds for any
	system, including non-chaotic (integrable, many-body-localized,...) systems and even systems without any notion of locality.
	
	In contrast, the second eigenstate-based temperature, $\beta_S$, is independent of the distance
	measure and reflects the physics of the eigenstates.  This contrast highlights that the partial
	trace operation is a crucial ingredient for the emergence of thermodynamics.
	We have shown that $\beta_S$ conforms increasingly to $\beta_C$ when the system size increases while
	keeping $L_A$ (subsystem size) fixed, and also while keeping the ratio $L_A/L$ fixed to some value
	smaller than $1/2$.  As $\beta_S$ depends on the chaotic (thermalizing) nature of the system and the physical content of the eigenstates, it does not match $\beta_C$ for random matrices, as shown in Appendix \ref{sec:Random},
	and generally shows deviant behavior for non-chaotic systems (Section \ref{sec:non_chaotic}).
	
	By asking how close $\rho^A$ can be to $\rho_C^A$, we have characterized the best temperature
	(typically different from the canonical temperature at finite sizes), and also the degree to which
	the system is thermal, e.g., through the value of the minimum distance $d_1$.
	The issues addressed in the investigation of $\beta_S$ are closely related to (in some sense the
	converse of) questions addressed in the ETH/thermalization literature, e.g., in
	Refs.~\cite{Muller2011, Garrison_PhysRevX.8.021026,YangLai__FFtypicality_PRB2015,
		TianYangWang_FFThermal_PRE2018, Dymarsky_Lashkari_Liu_PRE2018, LenarcicAltmanRosch_MBLinSolids_PRL2018,
		Lu_PhysRevE.99.032111, Murthy_PhysRevE.100.022131, 
		LenarcicRoschAltman_CriticalMBL_open_PRL2020,Burgdoerfer_Brezinova_arXiv2021, Fleckenstein_Bukov_PRB2021}.  Our results
	on size dependence confirms the intuition obtained from Refs.~\cite{Garrison_PhysRevX.8.021026,YangLai__FFtypicality_PRB2015,
		Lu_PhysRevE.99.032111, Murthy_PhysRevE.100.022131} that thermal behavior is best seen in the limit
	of $L_A/L\to0$.

	
	The present work raises a number of new questions.
	
	(1) The partial trace and minimization operations in the definition of $\beta_S$ render analytical
	treatments difficult.  Thus, it remains an open task to prove analytically that $\beta_S$ should be
	independent of $p$, or that it should approach $\beta_C$ in the large size limit.  The latter is
	consistent with the spirit of ETH, which is similarly difficult to prove, but is verified in a wide
	array of numerical studies \cite{rigol_thermalization_2008, Rigol_PRL2009, Rigol_PRA2009,
		Santos_Rigol_onset_PRE2010, RigolSantos_PRA10, BiroliKollathLauchli_PRL10,
		Roux_PRA2010_quantumquenches, Neunhahn_Marquardt_PRE2012, BrandinoKonikMussardo_PRB12,
		SteinigewegPrelovsek_PRE13,Rigol_Srednicki_PRL2013,
		SorgVidmarHeidrichMeisner_PRA14, SteinigewegGogolinGemmer_PRL2014,Kim_Huse_PRE2014,
		Beugeling_Moessner_Haque_PRE2014, Beugeling_Moessner_Haque_PRE2015, FratusSrednicki_PRE2015,Luitz_BarLev_PRL2016,
		MondainiSrednickiRigol_PRE16, Alessio_Rigol_AdvPhys2016, Mondaini_Rigol_2DIsing_PRE2017,
		Mori_Ikeda_Ueda_thermalizationreview_JPB2018, Sagawa_PRL2018,
		LeBlondVidmarRigol_PRE2019,JansenVidmarMeisner_PRB2019,
		Khaymovich_Haque_McClarty_PRL2019,
		HaqueMcClarty_SYKETH, Mierzejewski_Vidmar_PRL2020,
		Brenes_Goold_Rigol_XXZ_PRB2020,Brenes_Rigol_PRL2020,
		LeBlond_Rigol_PRE2020, Richter_Dymarsky_Gemmer_PRE2020,
		Prosen_ETH_DualUnitary_PRE2021, LiMaTan_JaynesCummingsHubbard_PhysicaScripta2021,
		HeidrichMeisner_Vidmar_PRB2021, Noh_PRE2021,
		Ueda_LocalRMT_PRL2021}.  Proving the $D^{-1/2}$ behavior of
	Fig.~\ref{fig:SubT_scaling}(B) also remains an open problem. 

	(2) The correspondence between $\beta_S$ and $\beta_C$ may break down when approaching non-chaotic
	regimes, such as near-integrability or many-body localization
	\cite{Oganesyan_Huse_localization_PRB2007, NandkishoreHuse_AnnuRev2015,
		Alet_Laflorencie_MBLreview2018}.  There is the possibility of scaling with different power-laws
	than those seen here, in analogy to the power-law ETH scaling displayed by integrable models
	\cite{Ziraldo_Santoro_PRB2013, Beugeling_Moessner_Haque_PRE2014, Beugeling_Moessner_Haque_PRE2015,
		Alba_PRB15, ArnabSenArnabDas_PRB16, HaqueMcClarty_SYKETH}. In Section \ref{sec:non_chaotic} we did observe the deviation of $\beta_{S}$ from $\beta_{C}$ as the system approached integrability, as one might have expected. A deeper investigation into the effects of integrability and localization is required.

	(3) A weak or even zero system-bath coupling is often considered the natural setting for discussing
	quantum thermalization \cite{Goldstein_PhysRevLett.96.050403, Riera_Gogolin_Eisert_PRL2012}.  In the
	present context, we did not consider it natural to modify $H_{AB}$, as we do not {\it a priori} have
	a system-bath separation, and the partition into $A$ and $B$ is arbitrary.  
	However, it would be interesting to explore the effect of varying $H_{AB}$.  
	For the exact limit of $H_{AB} = 0$,
	the reduced density matrix $\rho^{A}_{C} = \tr_B(\rho_{C})$ is just $e^{-\beta H_A}$, and the eigenstates of the full system decompose into tensor products of the eigenstates of the two subsystems.
	Thus the reduced eigenstate density matrix $\rho^{A}$ is simply $\outerprod{E_{j}^{A}}{E_{j}^{A}}$, where $\ket{E_j^A}$ are eigenstates of $H_A$.
	Thus, if one calculates the subsystem temperature $\beta_{S}$ for $H_{AB} = 0$, 
	the resulting temperature is actually the eigenstate temperature of the contributing eigenstate in $H_A$. Then using the result from Section \ref{sec:eigenstate_temperature}
	this temperature will in fact be $\beta_{C}/p$ of the subsystem $H_A$, as opposed to $\beta_{C}$ of the total system. One can still ask how the correspondence between $\beta_S$ and $\beta_C$ changes systematically in the $H_{AB}\to0$ limit.

	(4) In this work, we compared the eigenstate-based temperatures $\beta_{E}$ and $\beta_{S}$ to the canonical temperature $\beta_{C}$. The canonical temperature is widely used as a standard definition of temperature in the study of thermalization in many-body quantum systems. 
	In the study of statistical mechanics, a standard definition of temperature is the inverse of the derivative of entropy with respect to energy.
	The possibility of entanglement entropy being representative of the thermal entropy in the large system size limit is often discussed \cite{Garrison_PhysRevX.8.021026,Lu_PhysRevE.99.032111,Seki_Yunoki_PRResearch2020,Deutsch_Entropy_IOP2010,Deutsch_Sharma_PRE2013}. Ref.~\cite{Garrison_PhysRevX.8.021026} investigates the deviation of the entanglement entropy from a canonical entropy in a finite quantum system. One could consider the temperature arising from the entanglement entropy of eigenstates as another possible eigenstate based temperature.
	
	
	\begin{acknowledgments}
		We thank S.~Denisov, A.~Dymarsky and P.~A~McClarty for helpful
		discussions.  PCB thanks Maynooth University (National University of 
		Ireland, Maynooth) for funding provided via the John \& Pat Hume Scholarship. GN thanks the Irish Research Council for funding provided via the Government of Ireland Postgraduate Scholarship Program (Grant No. GOIPG/2019/58), and acknowledges financial support from the Deutsche Forschungsgemeinschaft (DFG) through SFB 1143 (project-id 247310070). The authors acknowledge the Irish Centre for High-End Computing (ICHEC) for the provision of computational facilities.
	\end{acknowledgments}
	
	
	\appendix
	
	
	\section*{Appendix}
	
	In the appendices, we provide additional numerical results:
	\begin{itemize}
		\item In Appendix \ref{sec:Random} we present numerical results for both the eigenstate and subsystem temperatures, using random matrices in place of a physical Hamiltonian. We illustrate the generality of the analytical result of $\beta_{C} = p\beta_{E}$, and also show how poorly $\beta_{S}$ and $\beta_{C}$ align for random matrices.
		\item In Appendix \ref{sec:subsys_data} we present further numerical data for the subsystem temperature, in particular, the result of varying the subsystem size in the staggered field model.
		\item In Appendix \ref{sec:altp}, we present results obtained using the Schatten 2-norm in place of the 1-norm for the subsystem temperature.
		\item In Appendix \ref{sec:betaC}, we compute the distance between $\rho^{A}$ and $\rho_{C}^{A}$ at the canonical temperature $\beta_{C}$.
	\end{itemize}
	
	
	\section{Random Matrix Results}\label{sec:Random}
	Here we present the results for both the eigenstate temperature $\beta_{E}$ and the subsystem temperature $\beta_{S}$ using a random, real, symmetric matrix in place of a physical Hamiltonian.
	
	\subsection{Eigenstate Temperature}
\begin{figure}[h]
		\includegraphics[width=0.313\linewidth]{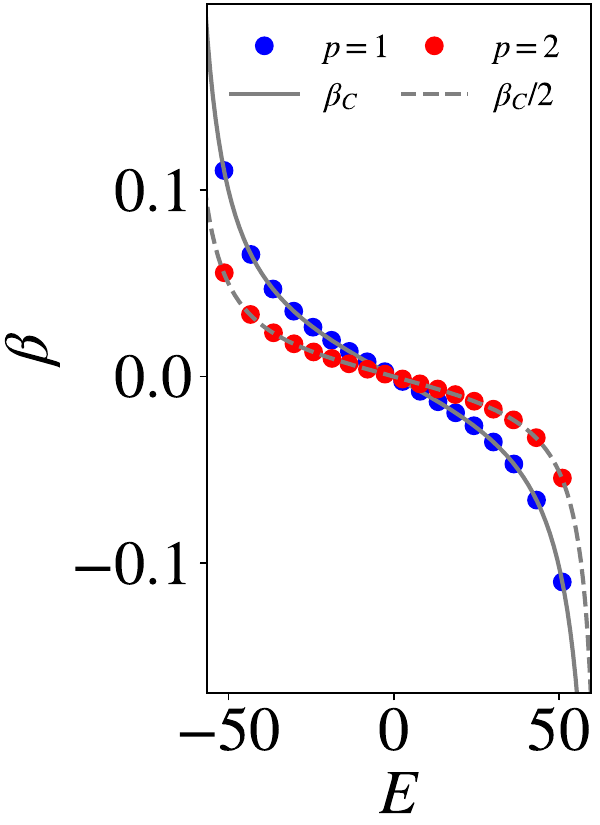}
		\includegraphics[width=0.351\linewidth]{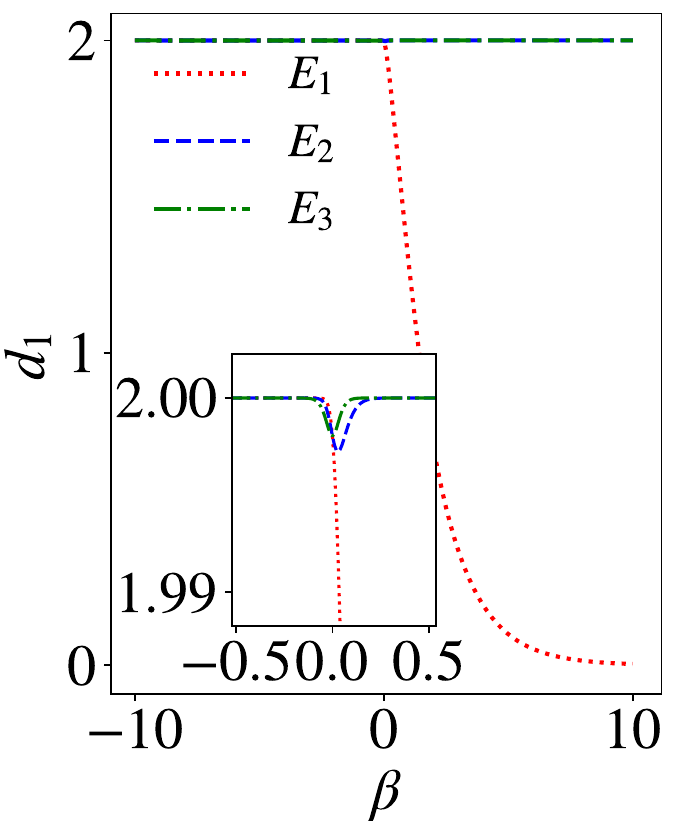}
		\includegraphics[width=0.313\linewidth]{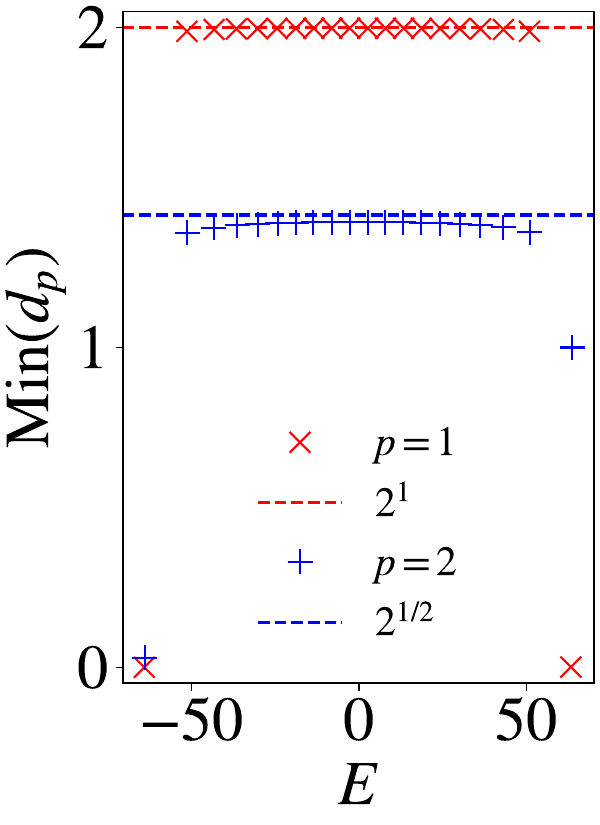}
		\caption{Eigenstate temperature results for random symmetric matrix with $D = 2^{10}$.
			\textbf{Left:} $\beta_E$ against energy, for 20 eigenstates which are equally spaced in energy across the spectrum, with curves showing $\beta_C/p$. (Highest/lowest state not visible.)  
			\textbf{Mid:} $d_1(\rho,\rho_{C})$ vs $\beta$ curve for ground state ($E_1$), mid-spectrum
			state ($E_3$), and $E_2$ in between the two.  
			\textbf{Right:} The minimum of $d_p(\rho,\rho_{C})$ plotted against
			eigenenergy, for the same
			eigenstates used in (A). 
		}
		\label{fig:rand_betaE}
	\end{figure}

	As previously demonstrated, $\beta_{C} = p\beta_{E}$ is a general mathematical result that will hold for any Hermitian matrix $H$. Here we illustrate this with a random matrix in Fig.$~$\ref{fig:rand_betaE}.
	
	\subsection{Subsystem Temperature}
\begin{figure}[h]
		\includegraphics[width=0.49\linewidth]{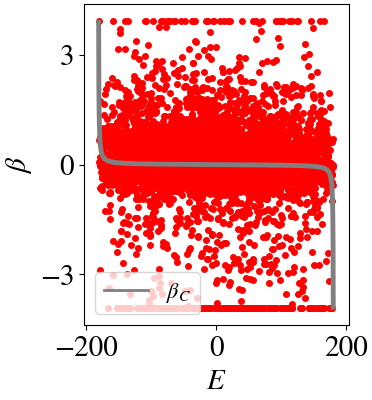}
		\includegraphics[width=0.49\linewidth]{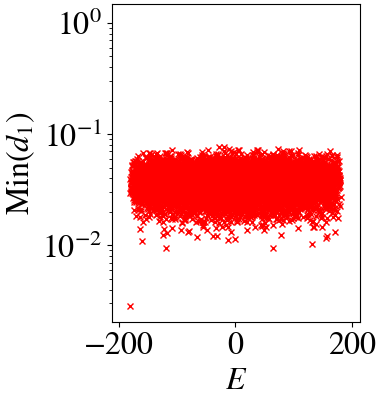}
		\caption{Subsystem temperature results for a random symmetric matrix with $D=2^{13}$. 
			\textbf{Left:} $\beta$ minimizing $d_1(\rho^{A},\rho_{C}^{A})$ ($\beta_{S}$) versus energy, plotted along side the canonical $\beta_{C}$ curve.
			\textbf{Right:} $\min(d_1(\rho^A,\rho^A_C))$ plotted versus energy.}
		\label{fig:rand_betaS}
	\end{figure}

	In the main text we found that $|\beta_{S} - \beta_{C}|\to 0$ when $L_A/L\to 0$, for the chaotic systems that we studied. Here we illustrate in Fig.$~$\ref{fig:rand_betaS} that this is not a generic result, showing how poorly the temperatures align for a random matrix.
	
	\section{Subsystem Temperature - Various subsystem sizes}\label{sec:subsys_data}
	
	\begin{figure}
		\includegraphics[width=\linewidth]{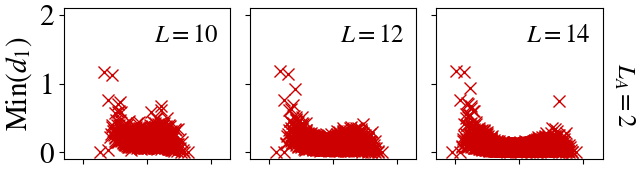}
		\includegraphics[width=\linewidth]{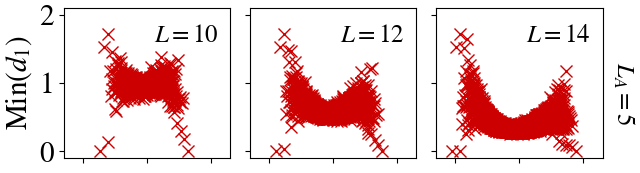}
		\includegraphics[width=\linewidth]{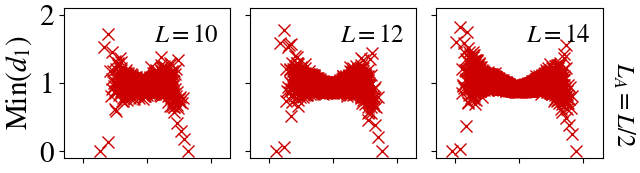}
		\includegraphics[width=\linewidth]{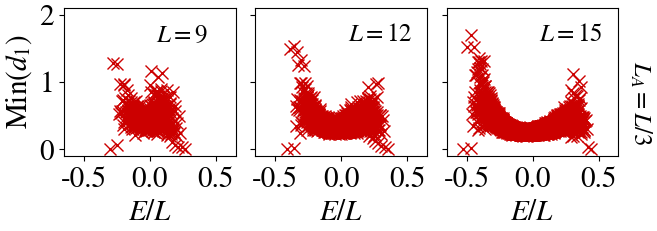}
		
		\caption{Subsystem temperature results for staggered field model with $h_x = h_z = 0.5$, $J=1$ and $\Delta=0.95$. $\min(d_1(\rho^A,\rho^A_C))$ plotted versus energy for the captioned $L$ and $L_A$.}
		\label{fig:eth_EvsX_rows}
	\end{figure}
	
	Here, we present the result of using different subsystem sizes when computing the subsystem temperature $\beta_{S}$, in various models.
	
	In Fig.$~$\ref{fig:eth_EvsX_rows} we show the resultant minimum $d_1$ when using different subsystem sizes for various system sizes. Illustrating again the decrease in average minimum distance as $L$ increases, but also showing that the average minimum distance increases with increasing $L_A$.
	
	\begin{figure}
		\includegraphics[width=\linewidth]{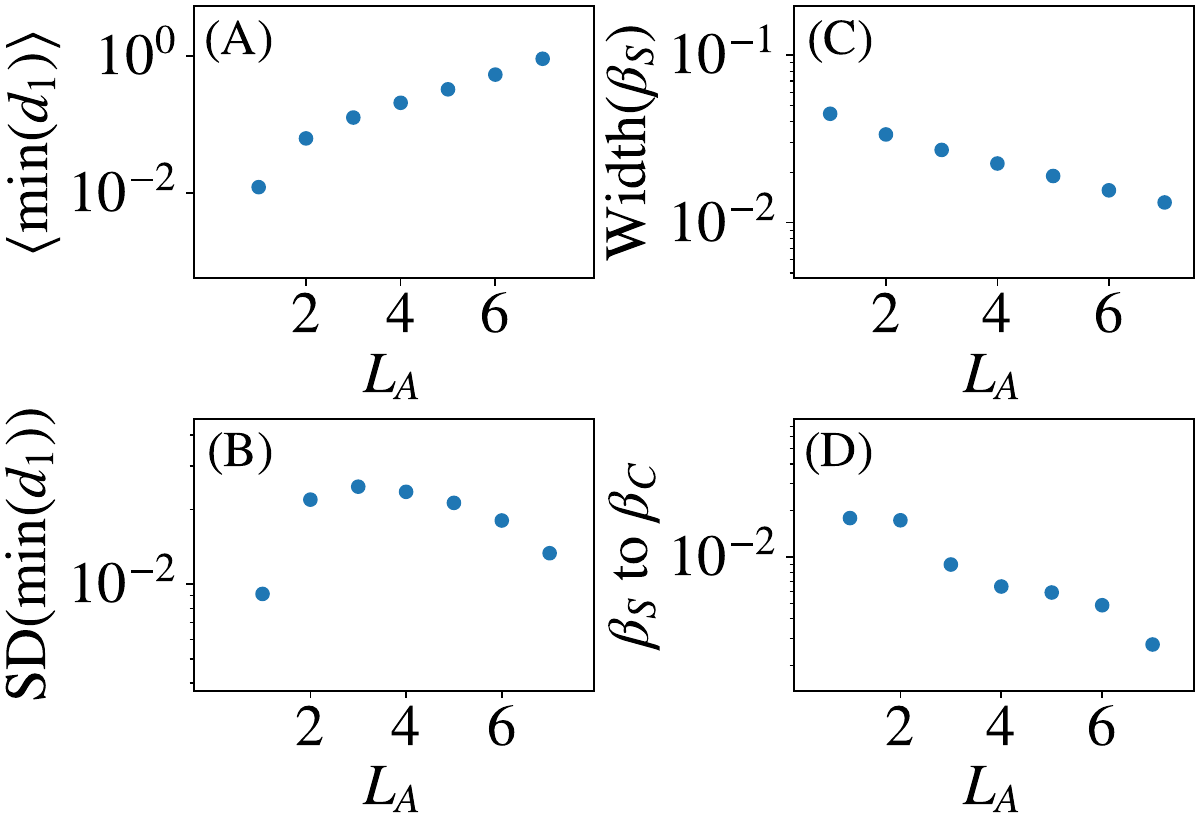}
		\caption{Subsystem temperature scaling with subsystem size results. Staggered field model with $h_x = h_z = 0.5$, $J=1$, $\Delta=0.95$ and $L=14$. \textbf{(A)} Mean value of $\min(d_1(\rho^{A},\rho_{C}^A))$, \textbf{(B)} Standard deviation of $\min(d_1(\rho^{A},\rho_{C}^A))$, \textbf{(C)} Width of $\beta_{S}$ data \textbf{(D)} RMS-distance between $\beta_{C}$ and linear fit to $\beta_{S}$, versus $L_A$. All quantities are calculated in the central 20\% of the spectrum.}
		\label{fig:La_scaling}
	\end{figure}

	In Fig.$~$\ref{fig:La_scaling} we show the explicit scaling of various quantities. We see in (A) that the average minimum of $d_1$ increases as $L_A$ increases, i.e., the two matrices become less alike. In (B) the standard deviation of the minima increases but then decreases again as $L_A$ approaches $L/2$. In (C) we see the width of $\beta_{S}$ decreased as $L_A$ increased, and similarly in (D) the distance between $\beta_{C}$ and $\beta_{S}$ decreased as $L_A$ increased.
	
	\begin{figure}
		\includegraphics[width=0.492\linewidth]{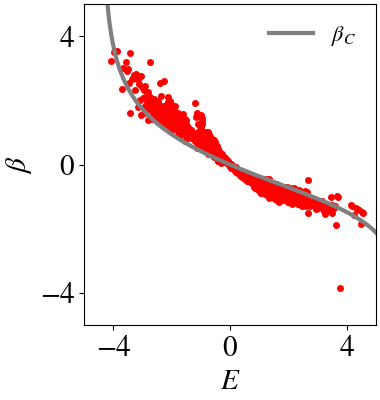}
		\includegraphics[width=0.492\linewidth]{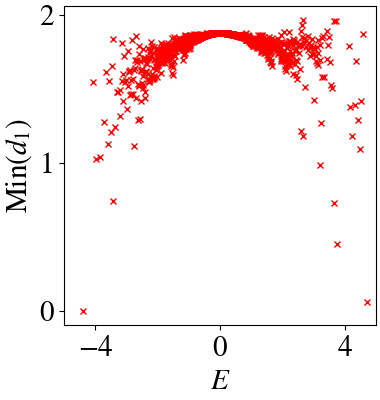}
		\caption{Subsystem temperature results for chaotic Ising model with $L=10$ and $L_A = 7$, $h_x = 0.75$ and $h_z = 0.5$. \textbf{Left:} $\beta_{S}$ vs. $E$ with canonical $\beta_{C}$ curve shown. \textbf{Right:} $\min(d_1(\rho^{A},\rho_{C}^A))$ versus energy.}	
		\label{fig:La_big}
	\end{figure}

	In the main text, we restricted our results to subsystems with $L_A < L/2$. Here we present an example of the result of using a subsystem with $L_A>L/2$. The minimum distance $\min(d_p(\rho^{A},\rho_{C}^{A}))$ continues the trend previously described of increasing as $L_A$ increases, and the variance of the values also decreased. However, the $\beta_{S}$ values appeared to cease to align with the $\beta_{C}$ curve, although the variance did continue to decrease. An example of the resultant $\beta_{C}$ for a subsystem greater than half the total system can be seen in Fig.$~$\ref{fig:La_big}. One can also see that the distance between the matrices is close to the maximum value.

	
	\section{Subsystem temperature with alternate p-distances}\label{sec:altp}
	
\begin{figure}
		\includegraphics[width=0.99\linewidth]{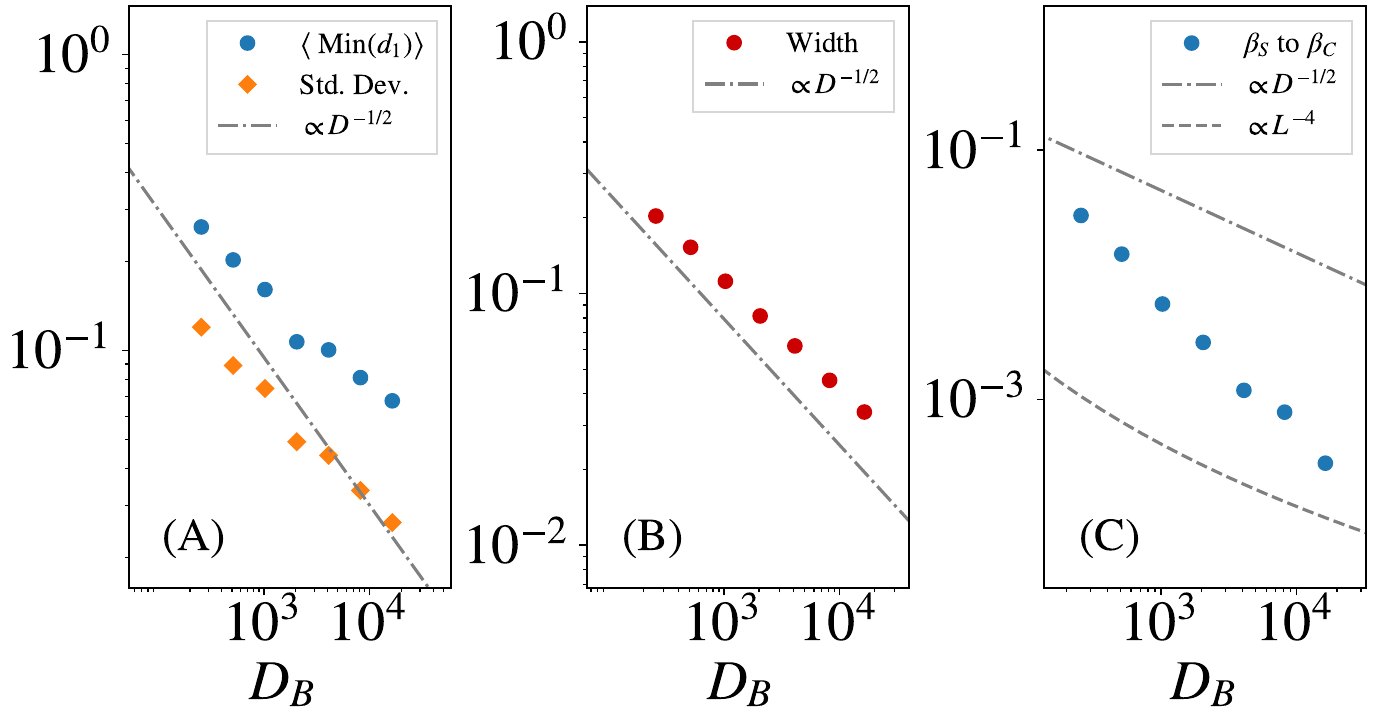}
		\caption{Subsystem temperature results using $p=2$ distance, for chaotic Ising model with $h_x = 0.75$, $h_z = 0.5$ and $L_A = 2$. Statistics from the central 20\% of the spectrum. 
			\textbf{(A)} Mean of $\min(d_2(\rho^A,\rho^A_C))$ and its
			standard deviation, vs. $D_B$.
			\textbf{(B)} Width of $\beta_{S}$ vs. $D_B$.
			\textbf{(C)} RMS-distance from the linear fit of $\beta_{S}$, to
			$\beta_{C}$ curve versus $D_B$.}
		\label{fig:p2eth_meanmin}
	\end{figure}

	In the main text, we showed there was an explicit $p$-distance dependence for the full eigenstate
	temperature, and stated that we found no similar dependence for the subsystem temperature.  In
	Fig.$~$\ref{fig:p2eth_meanmin} we show results for the Schatten 2-norm (Hilbert-Schmidt norm). The
	scaling results that we find are generally the same as those obtained for $p=1$. The only exception
	that we found was the $p=\infty$ distance (the operator norm), which resulted in a gap in
	$\beta_{S}$ around $\beta=0$.  Thus, in this case, $\beta_{S}$ was never close to $\beta_{C}$ where
	$\beta_{C}$ was near zero.
	
	\section{Distance at canonical temperature}\label{sec:betaC}
	
\begin{figure}
		\includegraphics[width=0.49\linewidth]{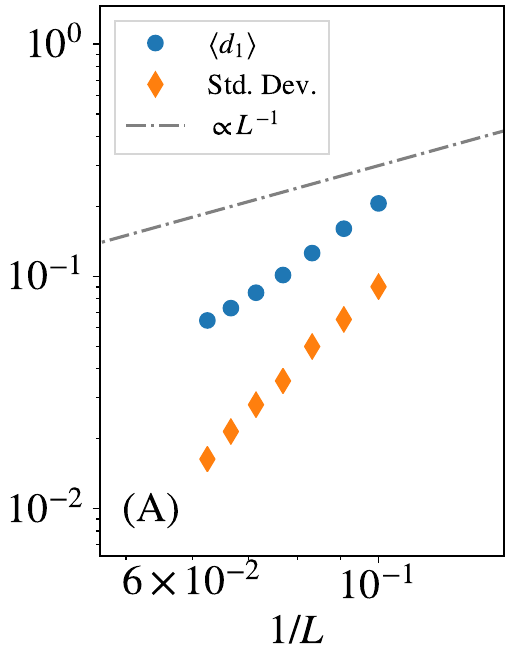}
		\includegraphics[width=0.49\linewidth]{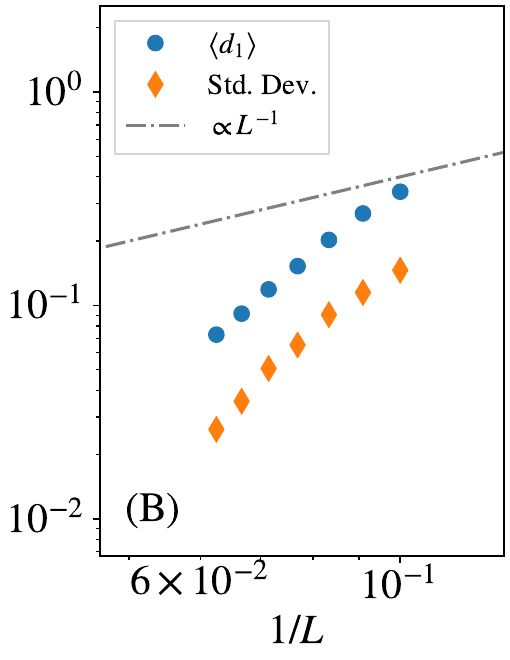}
		\caption{Distance $d_1$ at canonical temperature $\beta_{C}$, averaged over the central 20\% of the spectrum, 
			versus inverse system size $L$. We also show the mean standard deviation of the minima. With $L_A=2$ for (\textbf{A}) Staggered field  \textit{XXZ}-chain, with $h_z = h_x = 0.5$, and (\textbf{B}) Chaotic Ising model with $h_z = 0.5$ and $h_x = 0.75$.}
		\label{fig:eth_canbeta}
	\end{figure}

	In the main text, we minimized the distance between the reduced density matrix $\rho^{A} = \tr_{B}\outerprod{E_n}{E_n}$, and the reduced canonical matrix $\rho_C^A = \tr_{B}\exp(-\beta H)$, as a function of $\beta$, to obtain the subsystem temperature $\beta_{S}$. 
	One could instead ask how close the two matrices are at the canonical temperature $\beta_{C}$. In Fig.$~$\ref{fig:eth_canbeta} we show the resulting distances for the two chaotic models investigated in the main text. Alongside the data, we show a line proportional to the inverse system size $1/L$, which clearly illustrates that the distance between the matrices at $\beta_{C}$ decreases faster than $1/L$, for these particular systems at least.
	
	%

\end{document}